\documentclass[aps,pra,reprint,superscriptaddress,showpacs,showkeys]{revtex4-2}

\usepackage{caption}
\usepackage{comment}
\usepackage{appendix}

\usepackage{natbib}

\usepackage{amsmath, amssymb, amsfonts} %
\usepackage{amsthm}
\usepackage{xcolor}

\usepackage{float} 

\usepackage{subcaption}
\usepackage{graphicx} 
\usepackage{caption}
\usepackage{hyperref} 
\usepackage{xcolor} 
\usepackage{physics} 

\usepackage{tikz}
\usetikzlibrary{shapes.geometric, arrows}
\usepackage{circuitikz}
\usepackage{pgf-umlcd}

\tikzstyle{block} = [rectangle, rounded corners, minimum width=4cm, minimum height=1cm,text centered, draw=black, fill=blue!10]
\tikzstyle{arrow} = [thick,->,>=stealth]

\definecolor{dodgerblue}{RGB}{30, 144, 255} 

\hypersetup{
    colorlinks = true,
    linkcolor = dodgerblue,
    citecolor = dodgerblue,
    filecolor = dodgerblue,
    urlcolor = dodgerblue
}

\newtheorem{theorem}{Theorem}

\begin{document}

\title{Pontryagin Maximum Principle for Rydberg-blockaded state-to-state transfers:\\ 
A semi-analytic approach}

\author{Federico Astolfi}
\email{astolfi@unistra.fr}
\affiliation{University of Strasbourg and CNRS, CESQ and ISIS, 67000 Strasbourg, France}

\author{Sven Jandura}
\noaffiliation

\author{Guido Pupillo}
\email{pupillo@unistra.fr}
\affiliation{University of Strasbourg and CNRS, CESQ and ISIS, 67000 Strasbourg, France}

\date{\today}

\begin{abstract}
We study time-optimal state-to-state control for two- and multi-qubit operations motivated by neutral-atom quantum processors within the Rydberg blockade regime. Block-diagonalization of the Hamiltonian simplifies the dynamics and enables the application of a semi-analytic approach to the Pontryagin Maximum Principle to derive optimal laser controls. We provide a general formalism for $N$ qubits. For $N=2$ qubits, we classify normal and abnormal extremals, showcasing examples where abnormal solutions are either absent or suboptimal. For normal extremals, we establish a correspondence between the laser detuning from atomic transitions and the motion of a classical particle in a quartic potential, yielding a reduced, semi-analytic formulation of the control problem. Combining PMP-based insights with numerical optimization, our approach bridges analytic and computational methods for high-fidelity, time-optimal control.

\end{abstract}

\maketitle

\section{Introduction} \label{sec:introduzione}


Optimal control (OC) theory originates from the classical calculus of variations \cite{Santambrogio} and, since the pioneering work of Pontryagin and collaborators  \cite{Kaufman1964}, has become an established area of research in a variety of fields, including mathematics, physics, engineering and space science to name a few \cite{OC, Schattler, Koch2022, CasalinoMascolo2025EscapeL2}. 

In general terms, OC provides a framework for steering a process while minimizing a cost or maximizing a utility. This approach is particularly attractive in the field of quantum computing at the intersection of physics, computer science and mathematics, where a key goal is to execute dynamical operations -- e.g. to realize given unitary operations acting on the space of one or more qubits -- while optimizing a given fidelity function. The latter may quantify, for example, how accurately a state-to-state transfer can be realized in the presence of errors due to coupling to an external environment, imperfections in the control fields, etc. In the realm of quantum computing, quantum optimal control (QOC) has thus developed into a comprehensive framework encompassing diverse strategies for determining control parameters that maximize gate fidelities, minimize implementation times, and enhance robustness against common experimental imperfections \cite{Koch2022, Glaser2015SchrodingerCat}. In recent years, QOC methods have been implemented on a variety of platforms, such as neutral atoms \cite{Goerz2011PhaseGates, Muller2011EntanglingGates, Goerz2014RydbergRobust, Omran2019CatStates, Cui2017RydbergLattice, Smith2013CsControl, Anderson2015HighDimUnitary}, trapped ions \cite{Nebendahl2009OptimalControl, Choi2014MultimodeControl} and superconducting qubits \cite{Egger2014CZGates, Kelly2014RandomizedControl, HuangGoan2014FluxQubits, Werninghaus2014LeakageReduction}, and several different approaches have been developed. In general, it is found that low dimensional systems, e.g. single qubits, can allow for a fully analytic description of QOC dynamics \cite{DAlessandro, Sugny2, Sugny3, AlbertiniDAlessandro2015, FresseColson2025TimeOptimal, Romano}, while higher-dimensional cases are usually dealt with via numerical optimizations \cite{Sugny1, Sugny2, Khaneja2005, Reich2012, Werschnik2007}. Oftentimes, a combination of theoretical and numerical tools is employed to design optimal controls \cite{Sugny1, Sugny2, Sven, Pichler, Bergonzoni, PhysRevResearch.5.033052, PhysRevResearch.7.023063, PhysRevA.111.052625, Koch2022, petrosyan}.

Although these approaches may differ in their specific details and scope, they typically rely — at least to some extent — on the Pontryagin Maximum Principle (PMP) \cite{Kaufman1964, Sugny1, Sugny2, OC, Schattler, DAlessandro}. Similar to the Euler–Lagrange equations in calculus of variations, the PMP approach maximizes (or minimizes) a Hamiltonian along trajectories satisfying the state and adjoint equations. The PMP provides a widely used theoretical foundation to QOC, offering {\it necessary conditions} for the existence of optimal solutions. As such, it has been applied to find candidate solutions to a variety of quantum control problems \cite{Sven, Sugny3, AlbertiniDAlessandro2015, Romano, FresseColson2025TimeOptimal}. However, proving the actual {\it existence or uniqueness} of PMP candidate solutions  remains a largely unresolved problem, as no general method exists in QOC. Even when existence can be established, there is no universal strategy to actually find the solutions, and one ultimately has to rely on heuristics. 
Consequently, in practical applications, the PMP conditions are often treated as sufficient and solved numerically using methods such as gradient-based algorithms \cite{Sugny1,Khaneja2005}. It remains an interesting open question to provide analytical results -- and thus insights in QOC --  for the PMP problem in situations of physical importance. This includes state-to-state transfer and quantum gates in multi-qubit systems.

In this work, we apply the PMP to investigate the shape of global laser pulses for the time-optimal realization of state-to-state transfer of $N$ Rydberg atoms in the blockade regime \cite{Pichler, Sven, Gallagher1994RydbergAtoms}. The Rydberg blockade is a mechanism where strong interactions between nearby atoms excited to Rydberg states by a laser prevent more than one atom from being excited at the same time. In recent years, this mechanism has enabled a variety of breakthroughs in
quantum computing and quantum simulation with neutral atoms
\cite{MorgadoWhitlock, RevModPhys.82.2313, Saffman2016RydbergQC},
including the realization of high-fidelity multi-qubit $C_NZ$ gates ($N\ge 2$)
\cite{Sven, PhysRevLett.123.170503, Ma2023, Graham2019RydbergEntanglement, Pagano2022, Jandura2023OptimizingRydbergGates, Evered2023, Radnaev2025, Tao2025, PRXQuantum.6.020334, Tsai2024BenchmarkingAF}, the simulation of
exotic many-body systems 
\cite{Semeghini2021TopologicalSpinLiquid, Glatzle2014, Browaeys2020, Lahaye2009DipolarBosons, GonzalezCuadra2025, Bluvstein2021, Labuhn2016} and applications in quantum-enhanced metrology
\cite{cao2024multi}.
Here, (i) for the general case $N\ge2$, we demonstrate how to approach an analytic solution of the QOC problem of time-optimal state-to-state transfer and multi-qubit phase gates for $N$ Rydberg atoms in the blockade regime by leveraging the PMP statement. In accordance with literature, we classify candidate solutions between {\it normal} and  {\it abnormal extremals} \cite{Sugny1, OC, Schattler, Jandura2024}; (ii) for $N=2$, we prove that abnormal extremals, when they exist, correspond to a laser detuning that is constant in time; (iii) for $N=2$, we show that normal extremals are governed instead by a laser detuning that satisfies an Ordinary Differential Equation (ODE) with the same functional structure as the equation of motion of a particle evolving in a quartic potential. Therefore, generic simultaneous and time-optimal state-to-state transfers of two two-level systems implemented with Rydberg-blockaded atoms can be simply obtained by integrating the above ODE for the detuning. As examples of our approach, (iv) for $N=2$, we provide a complete analysis for two model cases: the simultaneous excitation of two two-level systems (\textbf{Case (i)}) and the implementation of a Controlled-Z (CZ) gate on two two-level systems initially in their ground states, up to single-qubit rotations (\textbf{Case (ii)}). In both cases, we  prove that abnormal extremals either do not exist (\textbf{Case (i)}) or are not time-optimal (\textbf{Case (ii)}). Normal extremals are then studied numerically by sampling the coefficients of the ODE governing the laser detuning, performing an optimization over all possible coefficients using a Broyden–Fletcher–Goldfarb–Shanno (BFGS) algorithm. Remarkably, we find perfect agreement with the optimal pulses found respectively in \cite{Pichler} and \cite{Sven}. The combination of analytic and numerical methods justifies referring to our approach as ``semi-analytic" and thus provides an interesting alternative to the fully-numerical solutions.

Our analysis shows how the Pontryagin Maximum Principle, through its standard classification of candidate extremals, provides a powerful framework for uncovering the structure of certain multi-qubit state-to-state transfer problems. For the Rydberg blockade under conditions of irradiation of the qubits by a single global laser pulse, this is fundamentally allowed by the reduction of the qubit Hamiltonian to a block-diagonal structure resulting in an ensemble of independent, generally non-equivalent effective two-level subsystems. 
The coexistence of different types of PMP extremals has concrete physical and control-theoretic implications, influencing both the trajectories and the qualitative behavior of optimal controls. Overall, this highlights the importance of incorporating the underlying mathematical theory to provide insights complementary to numerical optimization. 

The remainder of this paper is organized as follows.
Section~\ref{sec:rydberg} introduces the effective Rydberg blockade Hamiltonian for $N$  independent, generally non-equivalent two-level subsystems, together with its derivation for the phase-gate (Sec.~\ref{sec:gatesetting}) and cluster (Sec.~\ref{sec:clustersetting}) models.
Section~\ref{sec:OC} introduces the fundamentals of optimal control theory within the Pontryagin Maximum Principle approach and their physical context. State-to-state transfer in minimal time is formulated as a well-posed optimization problem.
Section~\ref{sec:semianalytic} presents our semi-analytic approach to time-optimal state-to-state transfer in two-level systems. Applying the Pontryagin Maximum Principle in the Rydberg blockade regime, we first present the formalism  for $N$ TLS in Sec.~\ref{sec:PMPtosols}. For the case of $N=2$ two-level systems, we classify normal and abnormal extremals: abnormal extremals require constant laser detuning and are shown to either non-exist or to be non–time-optimal [Sec.~\ref{sec:abnormalsols}]. Analysis of normal extremals reveals that, for generic transfers, the detuning follows the dynamics of a classical particle in a quartic potential — our main result [Sec.~\ref{sec:normalsols}]. Undetermined coefficients in the resulting equation are obtained numerically via BFGS optimization, yielding control profiles that coincide with Gradient Ascent Pulse Engineering (GRAPE) solutions and apply to arbitrary transfers.

Our approach can be applied to any state-to-state transfer and in Sec.~\ref{sec:normalsols}  we  provide additional examples of different trajectories beyond  \textbf{Cases (i)} and \textbf{(ii)}. Appendix~\ref{sec:appendixA} provides additional mathematical considerations not included in the main text.




\section{From Rydberg Blockade to Independent Two-level Systems} \label{sec:rydberg}

In this Section, we briefly recall two different models resulting in Rydberg blockade dynamics for atomic qubits: The \textbf{multi-qubit phase-gate} model and the \textbf{cluster} model. Both models permit a reformulation of their respective native Hamiltonians, resulting in effective dynamics equivalent to that of 
$N$ independent and inequivalent two-level systems (TLSs) with Hamiltonian
\begin{equation}
\label{TLS}
H_k(t) = \frac{\sqrt{k} \, \Omega(t)}{2} \bigg(e^{i\varphi(t)} |0\rangle{}_k {}_k\langle 1| + \text{h.c.}\bigg).
\end{equation}
In Eq.~\eqref{TLS}, $\Omega(t)$ and $\varphi(t)$ are respectively the laser amplitude and the laser phase that couple the internal ground state $|0\rangle{}_k$ and the excited state $|1\rangle{}_k$ of the $k-$th TLS, with $1\leq k \leq N$. Here and in the following, ``h.c." stands for the hermitian conjugate.

\begin{figure*}[t]  
    \centering
    \begin{subfigure}{0.48\textwidth}
        \centering
        \begin{tikzpicture}
            \draw[rounded corners=10pt] 
            (-1.2,-1.5) rectangle (7.2,1.8);

            \node at (0,-1) {$|00\rangle$};
            \draw [thick] (-0.5,-0.8) -- (0.5,-0.8);

            \node at (2,-1) {$|10\rangle$};
            \draw [thick] (1.5,-0.8) -- (2.5,-0.8);

            \node at (4,-1) {$|01\rangle$};
            \draw [thick] (3.5,-0.8) -- (4.5,-0.8);

            \node at (6,-1) {$|11\rangle$};
            \draw [thick] (5.5,-0.8) -- (6.5,-0.8);

            \node at (2,1.3) {$|r0\rangle$};
            \draw [thick] (1.5,1.) -- (2.5,1.);

            \node at (4,1.3) {$|0r\rangle$};
            \draw [thick] (3.5,1.) -- (4.5,1.);

            \node at (6,1.3) {$|W\rangle$};
            \draw [thick] (5.5,1.) -- (6.5,1.);

            \draw[<->, thick, violet!70] (2,-0.7) -- (2,0.9) node[midway,right] {\textcolor{violet!70}{$\Omega(t)$}};
            \draw[<->, thick, violet!70] (4,-0.7) -- (4,0.9) node[midway,right] {\textcolor{violet!70}{$\Omega(t)$}};
            \draw[<->, thick, violet!70] (6,-0.7) -- (6,0.9) node[midway,right] {\textcolor{violet!70}{$\sqrt{2} \Omega(t)$}};

            \node at (3,2.4) {\large \textbf{(a)}};
            
        \end{tikzpicture}
        \vspace{.75cm}
        \caption{\small Gate setting for $N=2$ Rydberg atoms.}
        \label{fig:schemagate}
    \end{subfigure}
    \hfill
    \begin{subfigure}{0.48\textwidth}
        \centering
        \begin{tikzpicture}
            \node[anchor=south west, inner sep=0] (img) at (0,0) 
                {\includegraphics[width=\textwidth]{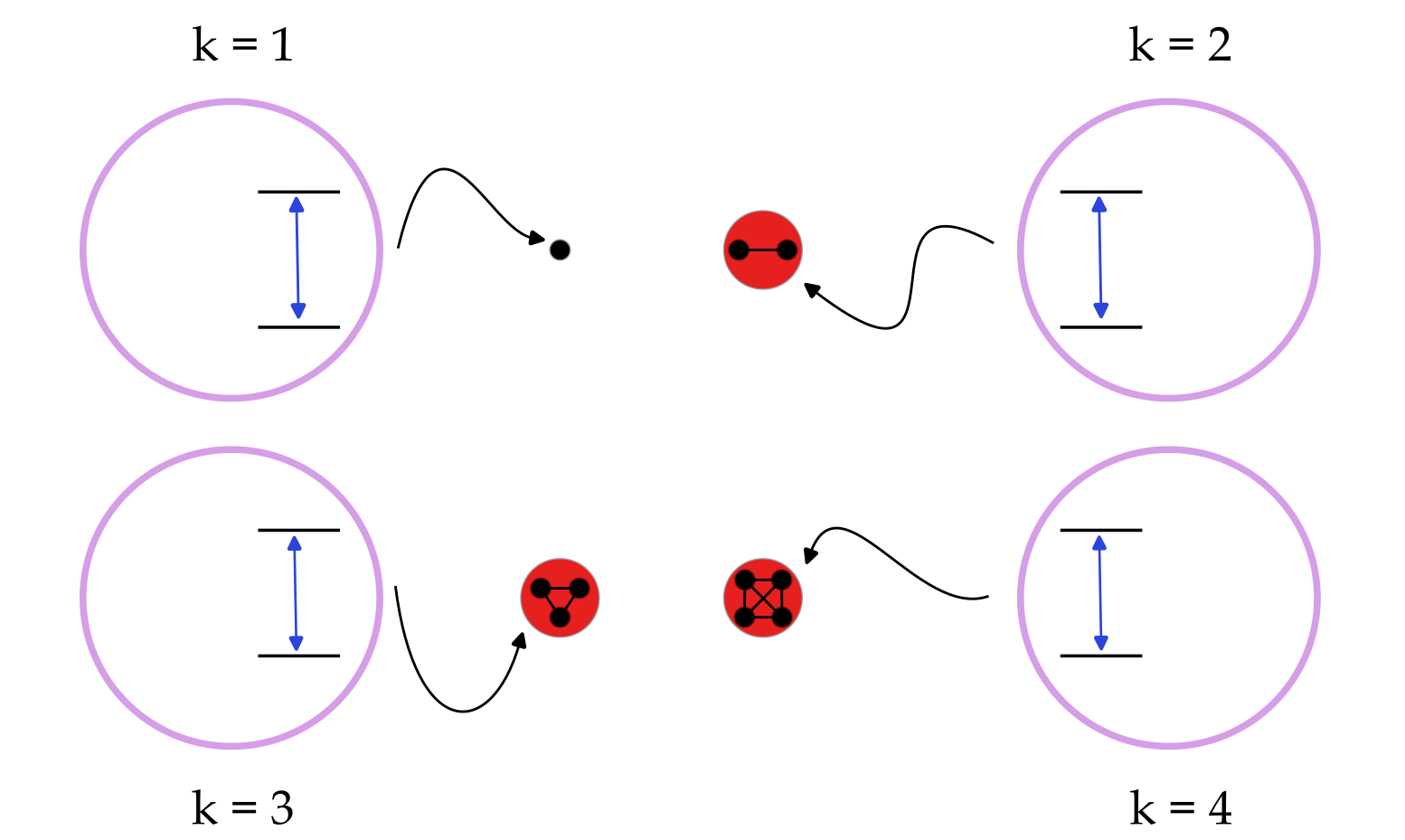}};
            
            \begin{scope}[shift={(0,0)}, font=\small]
                \node[blue!70] at (1.5, 3.6) {$\Omega(t)$};
                \node[blue!70] at (7.35, 3.6) {$\sqrt{2}\Omega(t)$};
                \node[blue!70] at (1.275, 1.5) {$\sqrt{3}\Omega(t)$};
                \node[blue!70] at (7.25, 1.5) {$2\Omega(t)$};
                \node at (1.1, 4) {$\ket{1}$};
                \node at (1.1, 3.2) {$\ket{0}$};
                \node at (1.1, 1.9) {$\ket{1}_3$};
                \node at (1.1, 1.1) {$\ket{0}_3$};
                \node at (7.6, 4) {$\ket{1}_2$};
                \node at (7.6, 3.2) {$\ket{0}_2$};
                \node at (7.6, 1.9) {$\ket{1}_4$};
                \node at (7.6, 1.1) {$\ket{0}_4$};
            \end{scope}

            \node at (4.425,4.65) {\large \textbf{(b)}};
            
        \end{tikzpicture}
        \caption{\small Cluster setting for $N=4$ clusters.}
        \label{fig:schema}
    \end{subfigure}
    \caption{\small Excitation processes for different Rydberg settings. (a) Scheme for the excitation process in the gate setting, for $N=2$ Rydberg atoms. The state $\ket{00}$ is not coupled to any excited state by the laser. The states $\ket{01}$ and $\ket{10}$ undergo a similar excitation process, by reaching the states $\ket{r0}$ and $\ket{0r}$ respectively. We remark that the laser amplitude $\Omega(t)$ has no coefficient in front, accounting for the fact that only one atoms is in the state $\ket{1}$. Finally, we see the excitation $\ket{11}\mapsto \ket{W}:=(\ket{r1}+\ket{1r})/\sqrt{2}$, that comes with amplitude $\sqrt{2}\Omega(t)$. (b) Scheme for the excitation process in the cluster setting, for $N=4$ clusters. We can see four different collections of Rydberg atoms, having an increasing number of atoms. Each one of the clusters behaves as a TLS, where the laser amplitude $\Omega(t)$ has to be rescaled by a factor $\sqrt{k}$, for $k=1,2,3,4.$}
    \label{fig:rydberg_schemes}
\end{figure*}

\subsection{Hamiltonian for multi-qubit controlled phase-gates via Rydberg blockade} \label{sec:gatesetting}

In the multi-qubit phase-gate model \cite{Sven, Pichler, PhysRevLett.123.170503}, we consider $N$ three-level atoms: a TLS is encoded in long-lived  $|0_j\rangle$ and  $|1_j\rangle$ states of atom $j$, while a Rydberg state $|r_j\rangle$ is used to mediate the interactions between different qubits via strong van der Waals (vdW) interactions $V_{\rm{vdW}}(R)\sim R^{-6}$ between Rydberg excited states -- with $R$ the interatomic distance \cite{MorgadoWhitlock}. The laser is tuned to the $|1_j\rangle\leftrightarrow |r_j \rangle$ transition, resulting in the following  Hamiltonian
\begin{equation}\label{rawpotential}
H = \frac{\Omega(t)}{2} \sum_j\bigg(e^{i\varphi(t)} |1_j\rangle \langle r_j| + \text{h.c.}\bigg) + \sum_{j<m} B_{jm} \hat{\mu}_j \hat{\mu}_m,
\end{equation}
with $B_{jm}$ the interaction
at a given distance $R$ and $\hat{\mu_j}=|r_j\rangle\langle r_j|$.
In experiments \cite{MorgadoWhitlock}, it is often the case that $B_{jm}=B\gg\Omega(t)$ for each $j<m$ and $t\ge0$. In the following, we thus focus on the so-called \textit{Rydberg blockade} limit $B/\Omega\rightarrow \infty$, which  prevents the population of states with more than one atom prepared in the Rydberg state $|r_j\rangle$. In such a regime, it is possible to block-diagonalize the Hamiltonian in Eq.~\eqref{rawpotential}  above, resulting in a decomposition into $N$ independent two-level systems, each governed by a lower-dimensional Schrödinger equation \cite{Sven}. While the state $\ket{0}^{\otimes N}$ is left unchanged by the Hamiltonian, every other state belonging to a given reduced computational basis -- i.e. block of the effective Rydberg blockaded Hamiltonian -- forms a closed two-level system coupling an effective ground state with an excited state containing one (and only one) Rydberg excitation. For example, for $N=3$, the states  $\ket{100}$, $\ket{010}$ and $\ket{001}$ are coupled respectively to  $\ket{r00}$, $\ket{0r0}$ and $\ket{00r}$ while the states $\ket{110}$, $\ket{101}$ and $\ket{011}$ are coupled respectively to $(\ket{r10}+\ket{1r0})/\sqrt{2}$, $(\ket{r01}+\ket{10r})/\sqrt{2}$ and $(\ket{0r1}+\ket{01r})/\sqrt{2}$. Any other excitation is forbidden.

More generally, for $N$ Rydberg atoms, the new effective $k$-th two-level system possesses a ground state $\ket{0}_k$, realized when the first $k$ atoms occupy the $\ket{1}$ state and the remaining $N-k$ atoms remain in $\ket{0}$.
For clarity of notation, we distinguish between subscripts written inside or outside the bra/ket symbols. When the subscript appears inside the parentheses --- as in $\ket{0_k}$, $\ket{1_k}$, and $\ket{r_k}$ --- it labels the different energy levels of the $k$-th atom. In contrast, a subscript written outside the bra/ket --- as in $\ket{0}_k$ and $\ket{1}_k$ --- denotes the effective ground and excited states defined above. In Eq.~\eqref{TLS}, the effective ground state $\ket{0}_k$ is coupled to its corresponding excited state $\ket{1}_k$ by the laser field.Consistent with the coupling $\ket{1} \leftrightarrow \ket{r}$ and within the Rydberg blockade regime, the laser induces an excitation on one — and only one — atom. As a result, the effective excited state $\ket{1}_k$ represents a single Rydberg excitation delocalized over multiple atoms. This procedure results in the decomposition of Eq.~\eqref{rawpotential}  into the $N$ independent Hamiltonians introduced in Eq.~\eqref{TLS}.
We note that all states containing the same number of atoms in state $\ket{1}$ evolve identically according to Eq.~\eqref{TLS}. Consequently, it is sufficient to select a single representative from each class of states to describe the dynamics of the entire system. 
For example, for $N=2$ atoms, we define $\ket{0}_1:=\ket{10}$, $\ket{1}_1:=\ket{r0}$, $\ket{0}_2:=\ket{11}$ and $\ket{1}_2:=(\ket{r1}+\ket{1r})/\sqrt{2}$. Block-diagonability of Eq.~\eqref{rawpotential} implies that the gate to be implemented is entirely determined by the evolution of the systems from the initial state $\ket{0}_1+\ket{0}_2$. 
Analogously, for $N=3$ atoms we can define $\ket{0}_1:=\ket{100}$, $\ket{1}_1:=\ket{r00}$, $\ket{0}_2:=\ket{110}$, $\ket{1}_2:=(\ket{r10}+\ket{1r0})/\sqrt{2}$, $\ket{0}_3:=\ket{111}$ and finally $\ket{1}_3:=(\ket{r11}+\ket{1r1}+\ket{11r})/\sqrt{3}$. The evolution of $\ket{0}_1+\ket{0}_2+\ket{0}_3$ again fully determines the evolution of the system.

\subsection{Cluster qubit Hamiltonian via Rydberg blockade} \label{sec:clustersetting}

In the cluster model \cite{Pichler, PhysRevLett.131.170601}, we consider arrays of two-level atoms with a ground state 
$\ket{0}$ and a Rydberg state $\ket{r}$, organized into multiple clusters—each consisting of neighboring atoms located within a blockade radius. This spatial configuration ensures that only a single Rydberg excitation can occur within each cluster, due to the Rydberg blockade effect. The clusters are sufficiently separated from one another so that inter-cluster interactions can be neglected.
In the presence of a laser tuned to the $|0_j\rangle\leftrightarrow |r_j \rangle
$ transition, the Hamiltonian reads
\begin{equation}\label{rawhamcluster}
H = \frac{\Omega(t)}{2} \sum_j\bigg(e^{i\varphi(t)} |0_j\rangle \langle r_j| + \text{h.c.}\bigg) + \sum_{\langle j, m \rangle} B_{jm} \hat{\mu}_j \hat{\mu}_m,
\end{equation} where the notation $\langle j,m\rangle$ stands for the second sum running only over atoms $j$ and $m$ in the same cluster and the rest of the notation is the same as in Eq.~\eqref{rawpotential}. Similar to the discussion above, we may block-diagonalize the Hamiltonian in Eq.~\eqref{rawhamcluster}, which results into Eq.~\eqref{TLS}. Here, a given cluster of $k$ atoms evolves in the space spanned by the effective ground and excited states defined  as  $\ket{0}_k:=\ket{0}^{\otimes k}$ and $\ket{1}_k:=\frac{1}{\sqrt{k}}\sum_{l=1}^k\ket{0}^{\otimes (l-1)}\otimes\ket{r}\otimes\ket{0}^{\otimes(k-l)}$, respectively. \\




\section{introduction to Optimal Control} \label{sec:OC}
Because the two-level systems described by the Hamiltonian in Eq.~\eqref{TLS} evolve independently, not every state-to-state transfer can be implemented. The accessible transfers are those that map states within the effective basis of each two-level system—where the $k$-th TLS is encoded in the Hilbert space spanned by the effective ground and excited states $\ket{0}_k$ and $\ket{1}_k$ (with the precise meaning of this notation depending on the physical models discussed in Subsecs.~\ref{sec:gatesetting} and \ref{sec:clustersetting})—onto the same basis, up to a global phase. This imposes a constraint on the class of target states reachable when $N\!\ge\!2$ independent and inequivalent TLSs are driven by Eq.~\eqref{TLS}. For instance, entangled states such as
$(\ket{0_1 0_2 \dots 0_N} + \ket{1_1 1_2 \dots 1_N})/\sqrt{2}$
are not accessible, since the transfer would fall outside the span of the effective basis. Nevertheless, the class of simultaneous transfers expressible as linear combinations of $\ket{0}_k$ and $\ket{1}_k$ is still rich, and we will consider two meaningful examples relevant to quantum computing.
\textbf{Case (i)} concerns the simultaneous excitation of two TLSs, in which both two-level systems are driven from their ground to their excited states \cite{Pichler}. \textbf{Case (ii)} is the generation of a \textit{CZ gate} (up to single gates) on the two TLSs in their ground state \cite{Sven}. The latter operation is defined by 

\begin{equation}
    \ket{xy}\mapsto(-1)^{xy}\ket{xy}.
\end{equation}
In the OC formalism, target states translate into target manifold. As for the two-TLSs examples above, \textbf{Case (i)} then translates into steering the effective ground states $(\ket{0}_1, \ket{0}_2)$ to  the manifold
\begin{equation}\label{torus}
\mathcal{N}=\big\{(e^{i\alpha}|1\rangle_1,e^{i\beta}|1\rangle_2):\alpha,\beta\in[0,2\pi]\big\}.
\end{equation}
Notice that this manifold is a {\it torus}. In contrast, \textbf{Case (ii)} yields an $1$-dimensional target manifold 
\begin{equation}\label{1dimtarget} \mathcal{N}=\big\{(e^{i\theta}|0\rangle_1,e^{i(2\theta+\pi)}|0\rangle_2):\theta\in[0,2\pi]\big\},\end{equation}
which must again be reached starting from $(\ket{0}_1, \ket{0}_2)$.

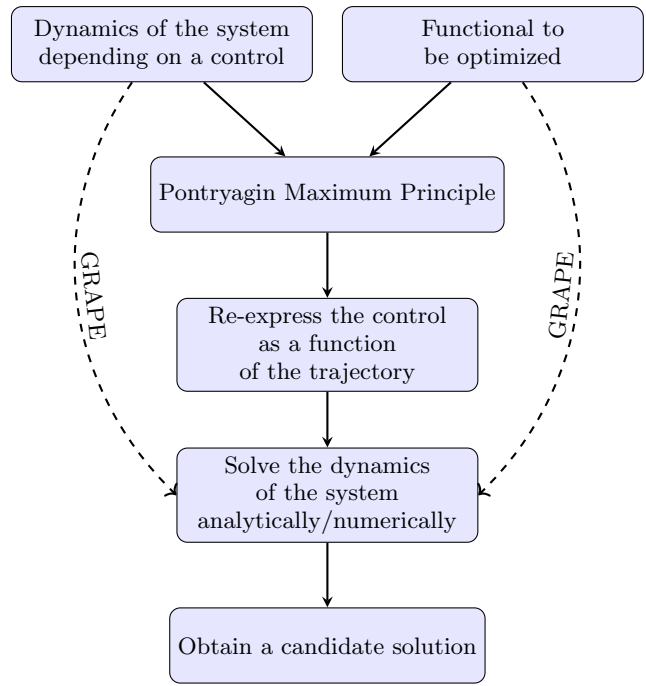
\begin{figure}[h]
    \centering
    \begin{circuitikz}
        \tikzstyle{block} = [draw, fill=blue!10, rounded corners, minimum width=4cm, minimum height=1cm, align=center]
        
        \node (A) at (-2.2,0) [block] {Dynamics of the system \\depending on a control };
        \node (F) at (2.2,0) [block] {Functional to\\be optimized};
        \node (B) at (0,-2) [block] {Pontryagin Maximum Principle};
        \node (C) at (0,-4) [block] {Re-express the control\\as a function\\of the trajectory};
        \node (D) at (0,-6) [block] {Solve the dynamics\\ of the system\\ analytically/numerically};
        \node (E) at (0,-8) [block] {Obtain a candidate solution};

        \draw[-stealth, thick] (A) -- (B);
        \draw[-stealth, thick] (F) -- (B);
        \draw[-stealth, thick] (B) -- (C);
        \draw[-stealth, thick] (C) -- (D);
        \draw[-stealth, thick] (D) -- (E);
        \draw[dashed, ->, bend right=40, thick] (A) to node[midway,above,sloped] {GRAPE} (D.west);
        \draw[dashed, ->, bend left=40, thick] (F) to node[midway,above,sloped] {GRAPE}(D.east);

    \end{circuitikz}
    \caption{Schematic approach to OC problems. A strategy to obtain a solution is to follow the solid lines in the scheme (cfr. Sec.~\ref{sec:semianalytic}). Therefore, the optimal candidate can be found either by solving the problem analytically or by leveraging numerical methods. An alternative is to follow the dashed lines in the scheme, i.e. to employ the GRAPE algorithm to devise approximated candidates \cite{Sugny1}.}
    \label{fig:pontryagin_diagram}
\end{figure}

This restatement will be useful below 
in search for time-optimal controls $\varphi(t)$ and $\Omega(t)$ driving the system to a \textit{given target state}, in the \textit{minimum time} possible. Overall, our approach to OC is summarized in the scheme of Fig. \ref{fig:pontryagin_diagram}.
\\

In the following, we begin by outlining the PMP, a fundamental mathematical result that yields candidate solutions to optimal control problems. We then formulate our model within the framework of the PMP and derive the corresponding necessary conditions. However, as said above, the PMP does not determine all the parameters required for a complete solution. Therefore, in Section~\ref{sec:semianalytic}, we address this indeterminacy using a combination of numerical and semi-analytic methods.

Say that $\mathcal{M}$ is a $m$-dimensional real manifold and consider a system of points $\mathbf{q}(t)=\{q_1(t),\dots,q_n(t)\}\subseteq \mathcal{M}$. Assume that the system is driven by
\begin{equation} \label{dynamics}
    \dot{\mathbf{q}}(t)=\mathbf{f}(\mathbf{q}(t),\mathbf{u}(t),t),
\end{equation}
where the function $\mathbf{u}(t)=(u_1(t),\dots,u_k(t)):\mathbb{R}\to U \subseteq \mathbb{R}^k$, almost everywhere continuous, is called the \textbf{control} of the system and the function $\mathbf{f}=(f_1,\dots,f_n)$ is sufficiently smooth \cite{OC,Schattler}. Any control that satisfies Eq.~\eqref{dynamics} is referred to as an \textbf{admissible control}.
The concept of system control has an intuitive interpretation: given a set of possible control inputs, the goal is to select the one that optimizes a specific quantity associated with the system, either by minimizing or maximizing it. \\
As stated in the end of Section~\ref{sec:rydberg}, we want to address a \textit{time-optimal} problem, namely we wish to steer the dynamics of Eq.~\eqref{dynamics} in a way that some given target state is reached and the overall time is minimized. This is summarized in the following conditions
\begin{equation} \label{OCproblem}
    \left\{
    \begin{aligned}
        &\min_{\mathbf{u}\in U} \, T, \\
        &\dot{\mathbf{q}} = \mathbf{f}(\mathbf{q}(t), \mathbf{u}(t), t), \\
        &\mathbf{q}(0) = \mathbf{q}_0, \\
        &\mathbf{q}(T) \in \mathcal{N}, \\
        &\mathbf{u}(t) \in U,
    \end{aligned}
    \right.
\end{equation}
where $\mathbf{q}_0$ is a set of $n$ points within $\mathcal{M}$ and $\mathcal{N}\subseteq \mathcal{M}$ is a submanifold, referred to as \textbf{target submanifold}.

We apply the PMP in order to derive the necessary conditions for optimality of candidate solutions to Eq.~\eqref{OCproblem}. This provides a semi-analytic framework to analyze such candidates. The core idea is to exploit the structure imposed by the PMP to express the control $\mathbf{u}$ in terms of the trajectory $\mathbf{q}$. Substituting this control back into the dynamical equations allows us to investigate the resulting system either analytically or numerically. The PMP is stated as follows.

\begin{theorem}[Pontryagin Maximum Principle] \label{teoPMP}
Suppose that $\mathbf{u}^*(t)$ is an admissible control relative to 
 Eq. \eqref{OCproblem} with respect to a trajectory $\mathbf{q}^*(t)$. Then it exists a pair of \textbf{costates} $(\lambda^*_0,\lambda^*)$ with the properties that 
\begin{itemize}
    \item $\lambda_0^* \le 0$ is a constant;
    \item $\lambda^*:[0,T]\to T_{\mathbf{q}^*(t)}\mathcal{M}$ is continuous.
\end{itemize}
Above, the notation $T_{\mathbf{q}^*(t)}\mathcal{M}$ stands for the tangent space to $\mathcal{M}$ at the point $\mathbf{q}^*(t)\in \mathcal{M}$ \footnote{Recall that $T_{\mathbf{q}^*(t)}\mathcal{M}\simeq\mathbb{R}^{\text{dim}\mathcal{M}}$.}.

Define the \textbf{pseudo-Hamiltonian} as
\[\mathcal{H}(t,\mathbf{q},\mathbf{u},\lambda_0,\lambda)=
\langle \lambda, \dot{\mathbf{q}} \rangle_{T_{\mathbf{q}^*(t)}\mathcal{M}} + \lambda_0,\]
where $\langle \cdot , \cdot \rangle_{\mathcal{M}}$ is the scalar product defined on the vector space $T_{\mathbf{q}^*(t)}\mathcal{M}$. The costates are such that
\begin{enumerate}
    \item $(\lambda_0^*,\lambda^*)\not = (0,\mathbf{0});$
    \item the following \textbf{Adjoint Equations} hold:\[\dot\lambda^*=-\frac{\partial\mathcal{H}}{\partial\mathbf{q}^*};\]
    \item the following \textbf{Maximum Principle} holds:
    \[\mathcal{H}(t,\mathbf{q}^*,\mathbf{u}^*,\lambda^*_0,\lambda^*)=\max_{\mathbf{u}\in U}\mathcal{H}(t,\mathbf{q}^*,\mathbf{u},\lambda_0^*,\lambda^*);\]
    \item the pseudo-Hamiltonian vanishes along the optimal trajectory: \[\mathcal{H}(t,\mathbf{q}^*,\mathbf{u}^*,\lambda^*_0,\lambda^*)=0;\]
    \item the following \textbf{Transversality Conditions} hold:
    \[\langle \lambda^*(t), v\rangle_{T_{\mathbf{q}^*(T)}\mathcal{N}}=0\] for any $v\in T_{\mathbf{q}^*(T)}N$, i.e. the costates $\lambda^*$ relative to some candidate trajectory are perpendicular at the final time $T$ to the target manifold $\mathcal{N}$.
\end{enumerate}
\end{theorem}
\noindent It is standard \cite{Sugny1, OC, Schattler, Jandura2024} to define \textbf{extremal} a trajectory $\mathbf{q}^*$ satisfying the necessary conditions for optimality as stated above. Furthermore, extremals such that  $\lambda_0^*<0$ are referred to as \textbf{normal} extremals. Instead, $\lambda^*_0=0$ leads to \textbf{abnormal} extremals. Clearly, costates can be renormalized when the extremals are normal \cite{OC}; as a consequence, without loss of generality we will assume that $\lambda^*_0$ is $0$ for abnormal extremals and $-1$ for normal extremals.
To shed light on the intuition behind the costates $\lambda_0^*$ and $\lambda^*$, we can think of them as analogous to  Lagrangian multipliers that play a relevant role in multi-variable calculus.\\

In Section \ref{sec:rydberg}, we defined the Rydberg blockade dynamics in order to devise a state-to-state transfer. In  QOC, the information regarding the target state (to be reached up to global phase factors) is stored inside the target manifold of the problem. In the following, by choosing a specific target manifold, we implement specific simultaneous state-to-state transfers of $N$ TLSs. 
To simplify notation, asterisks will be dropped whenever referring to extremals, their controls and their costates. We refer to Appendix~\ref{sec:appendixA} for the details on how the transition is made from the complex quantum mechanical framework to the real set-up formulation of the PMP.

Say $\ket{\psi_1(t)},\dots, \ket{\psi_N(t)}$ are $N$ TLSs in their respective effective ground states $\ket{0}_k$, whose dynamics is governed by Eq.~\eqref{TLS}.
As a consequence, Eq.~\eqref{OCproblem} translates into

\begin{equation} \label{QOC}
    \left\{
    \begin{aligned}
        & \min_{\varphi, \Omega} \, T, \\
        & \dot{\ket{\psi_k}} = -i H_k \ket{\psi_k}, \qquad k = 1,\dots,N, \\
        & \ket{\psi_k(0)} = \ket{0}_k, \qquad k = 1,\dots,N, \\
        & \big( \ket{\psi_1(T)}, \dots, \ket{\psi_N(T)} \big) \in \mathcal{N}, \\
        & \varphi(t) \in \mathbb{R}, \\
        & \Omega(t) \in [0, \Omega_{\max}].
    \end{aligned}
    \right.
\end{equation}
The target manifold $\mathcal{N}$ depends on the given target state and $\Omega_{\max}$ describes the maximum frequency achievable in a given experiment. By comparison with Eq.~\eqref{OCproblem}, we may recognize the trajectory $\mathbf{q}(t)=(\ket{\psi(t)}_1,\dots,\ket{\psi(t)}_N)$, the controls $\mathbf{u}(t)=(\varphi(t),\Omega(t))$ and function ruling the dynamics $\mathbf{f}(t)=-i(H_1(t)\ket{\psi(t)}_1,\dots,H_N(t)\ket{\psi(t)}_N)$. This identification completes the mapping of the desired control model for the dynamics of Eq. \eqref{TLS} into the framework of optimal control and PMP. \\

We note that time optimality dictates that $\Omega(t)=\Omega_{\max}$ at each time \cite{Sven}. In the following, without loss of generality, we set $\Omega_{\max}=1$ and solve the model as depending on only one control: the laser phase $\varphi(t)$. Since $\varphi$ is now the only unknown of the laser pulse to be determined, we will refer interchangeably to $\varphi$ either as laser phase, laser pulse or, simply, control.
Finally, we recall that at the beginning of Sec.~\ref{sec:OC} we introduced Eq.~(\ref{torus}) as the target manifold for the state-to-state transfer described in \textbf{Case (i)}, while \textbf{Case (ii)} was expressed by Eq.~(\ref{1dimtarget}). These two equations provide concrete examples of the target manifold $\mathcal{N}$ appearing in Eq.~\eqref{QOC}, and their definitions will be used to solve the corresponding models in a semi-analytic fashion in Sec.~\ref{sec:semianalytic}. In the next section, we first unravel the PMP as applied in Eq.~\eqref{QOC} to a generic system of $N$ TLSs governed by Eq.~\eqref{TLS}, and then discuss the extremals arising for the case of $N=2$ TLSs. We prove as the main result that for $N=2$ TLSs, the detuning of the laser pulse $\Delta(t)=\dot\varphi(t)$ driving the dynamics in minimal time must satisfy the ODE Eq.~\eqref{eqdiffpotential}. In order to explicitly obtain a solution, we must numerically optimize over the coefficients of this ODE.




\section{The semi-analytic approach} \label{sec:semianalytic}

In this Section, we apply the PMP from Theorem~\ref{teoPMP} to the QOC problem in Eq.~\eqref{QOC}. We first present the general formulation for $N$ TLSs and then focus on the low-dimensional case $N=2$. This simpler setting allows us to show that abnormal extremals have constant detuning; in particular, we prove that they are absent in \textbf{Case (i)} and strictly slower than the optimal solution in \textbf{Case (ii)} \cite{Sven, Jandura2024}. Normal extremals for $N=2$ form the core of our analysis: we find that, for any state-to-state transfer, the detuning evolves as a particle in a quartic potential, and we obtain the corresponding solutions by numerically integrating the resulting ODE. We numerically integrate Eq.~\eqref{eqdiffpotential} and confront the extremals with the GRAPE solution, and we see that they match. As such, our method is alternative to GRAPE and provides an interesting example on how rich is the information that can be extrapolated from the PMP, when applied to the control of multi-qubit quantum systems.

\subsection{From PMP to candidate extremals} \label{sec:PMPtosols}
We re-express the Hamiltonian Eq.~\eqref{TLS} as \begin{equation}\label{hk2}
    H_k = \frac{\sqrt{k}}{2}\big(\cos\varphi \sigma_x - \sin\varphi \sigma_y \big),
\end{equation}
where $\sigma_x, \sigma_y$ are Pauli matrices and $\Omega(t)=\Omega_{\max}=1$ for all $t$ (see above). 

Since the PMP is formulated for real manifolds, the quantum framework require to adapt accordingly the notions of inner product and derivatives. To this end, the real part of the complex hermitian product and the Wirtinger derivatives respectively are employed. See App.~\ref{sec:appendixA} for further details.\\
We then apply the PMP to Eq.~\eqref{QOC}, with $H_k$ from Eq.~\eqref{hk2}.
Consider a solution $(|\psi_1(t)\rangle,\dots,|\psi_N(t)\rangle,\varphi(t))$, where each TLS $\ket{\psi_k(t)}$ belongs to the Bloch sphere for all $t$. The PMP guarantees the existence of costates $\chi_0 \in \{0,-1\}$ and $|\chi_1(t)\rangle,\dots,|\chi_N(t)\rangle$ satisfying
\[
\Re\langle \chi_k(t)\,|\,\psi_k(t)\rangle = 0 \qquad \text{for all } t \text{ and each } k \ge 1.
\]
This orthogonality condition arises because the costates $|\chi_k(t)\rangle$ lie in the tangent space of the Bloch sphere, and tangent vectors have zero scalar product with points on the sphere.
Moreover, from the PMP Theorem \ref{teoPMP} we introduce the pseudo-Hamiltonian 

\begin{align} \label{zanza}
\mathcal{H}\big(t,|\psi\rangle,\varphi,\chi_0,|\chi\rangle\big)
&=\Im\langle\chi|H(\varphi)|\psi\rangle+\chi_0 \notag \\
&= \sum_k \Im \langle \chi_k | H_k | \psi_k \rangle + \chi_0 \notag \\
&= \frac{1}{2} \cos \varphi \left( \sum_k \sqrt{k} \Im \langle \chi_k | \sigma_x | \psi_k \rangle \right) \notag \\
&\quad - \frac{1}{2} \sin \varphi \left( \sum_k \sqrt{k} \Im \langle \chi_k | \sigma_y | \psi_k \rangle \right) + \chi_0. 
\end{align}

We now unfold one by one the statements from the PMP Theorem \ref{teoPMP}:
\begin{enumerate}
    \item $(\chi_0,|\chi(t)\rangle)\not = (0,\mathbf{0})$.
    \item The costates satisfy \[|\dot\chi_k\rangle = -\frac{\partial\mathcal{H}}{\partial|\psi_k\rangle}=-iH|\chi_k\rangle,\] i.e. they are steered by the same time dependent Schrödinger equation of $|\psi_k\rangle$.
    \item The Maximum principle \[\mathcal{H}\big(|\psi(t)\rangle, |\chi(t)\rangle, \chi_0, \varphi(t)\big) = \max_{\xi} \mathcal{H}\big(\psi(t), \chi(t), \chi_0, \xi\big)
\]
holds for every time $t$. In order to solve for $\varphi$, we impose $\frac{\text{d}\mathcal{H}}{\text{d}\varphi}=0$ and find that
\begin{equation}\label{cos}
    \cos \varphi = \frac{\sum_k \sqrt{k} \Im \langle \chi_k | \sigma_x | \psi_k \rangle}
{\omega}
\end{equation}
and 
\begin{equation}\label{sin}
    \sin \varphi = - \frac{\sum_k \sqrt{k} \Im \langle \chi_k | \sigma_y | \psi_k \rangle}
{\omega},
\end{equation}with 
\begin{equation}\label{omega}
\small
\omega = \sqrt{\left( \sum_k \sqrt{k} \Im \langle \chi_k | \sigma_x | \psi_k \rangle \right)^2 + 
\left( \sum_k \sqrt{k} \Im \langle \chi_k | \sigma_y | \psi_k \rangle \right)^2}
\end{equation}
whenever $\omega\not=0$ (when $\omega = 0$, the control $\varphi$ cannot be expressed as a function of $\ket{\psi_k}$ and $\ket{\chi_k}$ in this way, see Condition 4 below).
We then substitute Eqs.~\eqref{cos} and~\eqref{sin} for the control $\varphi$ into Eq.~\eqref{zanza} and obtain
\begin{align}\label{ham100}
\mathcal{H} 
&= \frac{1}{2} \left[
\left( \sum_k \sqrt{k} \Im \langle \chi_k | \sigma_x | \psi_k \rangle \right)^2  \right. \\
& \quad \left. + \left( \sum_k \sqrt{k} \Im \langle \chi_k | \sigma_y | \psi_k \rangle \right)^2
\right]^{1/2} + \chi_0. \notag
\end{align}

\item The pseudo-Hamiltonian in Eq. \eqref{ham100} vanishes over the extremals and the associated costates: 
\begin{equation}\label{H=0}
\mathcal{H}\big(|\psi(t)\rangle, |\chi(t)\rangle, \chi_0, \varphi(t)\big) =0
\end{equation}
at each time $t$. As mentioned previously in Sec.~\ref{sec:OC}, we define abnormal extremal an extremal for which $\chi_0=0$ and normal extremal one such that $\chi_0\not=0$. Therefore, normal extremals (Sec.~\ref{sec:normalsols}) lead to $\omega\not=0$: the aforementioned description of the control $\varphi$ can be employed (see Condition 3 above). Conversely, abnormal extremals (Subsec.~\ref{sec:abnormalsols}) lead to $\omega=0$ in Eq.~\eqref{omega} and thus we cannot express the control by means of the states $\ket{\psi_k}$ and the costates $\ket{\chi_k}$ as in Eqs.~\eqref{cos} and~\eqref{sin}.

\item The projection of the costates over the tangent space of the target manifold $\mathcal{N}$ is null at the final time $T$. As an explicit example, we consider 
\begin{equation}\label{ntorus}
\mathcal{N}=\big\{(e^{i\alpha_k}|1\rangle_k)_k:\alpha_k\in[0,2\pi]\big\},
\end{equation}
which is the $N$-TLSs generalization of the target manifold~\eqref{torus}  of \textbf{Case (i)} with $N=2$. The respective generalization for $N$-TLSs of \textbf{Case (ii)} is similar and henceforth we omit here its explicit computation.
Condition 5 from the PMP expressed in Theorem \ref{teoPMP} refers to the components of costates $|\chi(T)\rangle=\big(|\chi_1(T)\rangle,\dots,|\chi_N(T)\rangle\big)$ being zero along the generators of the tangent space of $\mathcal{N}$ in $|\psi(T)\rangle=\big(|\psi_1(T),\dots,|\psi_N(T)\rangle\big)$. We compute the first generator of the tangent space as \[
\frac{\partial |\psi(T)\rangle}{\partial \alpha_1} = i\big(e^{i\alpha_1}|1\rangle_1, 0,\dots,0\big)=i\big(|\chi_1(T)\rangle,0,\dots,0\big).\]
More generally, for each $k=1,\dots,N$, \[\frac{\partial |\psi(T)\rangle}{\partial \alpha_k} =i\big(0,\dots,0,|\chi_k(T)\rangle,0,\dots,0\big).\]
By imposing orthogonality with $\mathcal{N}$ \[\ket{\chi(T)}
\perp 
\mathcal{N},
\]
we obtain
\[0=\Re\langle\chi_k(T)|i\psi_k(T)\rangle=-\Im\langle\chi_k(T)|\psi_k(T)\rangle\] for each TLS. Due to Condition 2, $|\chi_k\rangle$ does satisfy the same time dependent Schrödinger equation  of $|\psi_k\rangle$ and thus their hermitian product $\langle\chi_k|\psi_k\rangle$ is preserved in time. As a consequence, we understand that \begin{equation}\label{Imaginary}\Im\langle\chi_k(t)|\psi_k(t)\rangle=0\end{equation} for all times $t$.

\end{enumerate}

In Subsecs.~\ref{sec:abnormalsols} and~\ref{sec:normalsols}, it will be useful to consider the following $3$-dimensional vectors
\[
\mathbf{v}_k = \left(v_k^{(x)}, v_k^{(y)}, v_k^{(z)}\right)
\]
given by 
\begin{equation}\label{description}
v_k^{(\mu)} = \text{Im} \langle \chi_k | \sigma_\mu | \psi_k \rangle, \quad \text{for } \mu \in \{x, y, z\}.
\end{equation}
We take the time derivative and obtain
\begin{equation}
\dot{v}_k^{(\mu)} = \text{Im}\big(i \langle \chi_k | [H_k, \sigma_\mu] | \psi_k \rangle\big), 
\end{equation}
which leads to the system of differential equations
\begin{align}\label{humongous}
\dot{\mathbf{v}}_k &= \sqrt{k} 
\begin{pmatrix}
\cos \varphi \\
-\sin \varphi \\
0
\end{pmatrix}
\times \mathbf{v}_k \notag \\
&= \sqrt{k} 
\begin{pmatrix}
-\sin \varphi v_k^{(z)} \\
-\cos \varphi v_k^{(z)} \\
\cos \varphi v_k^{(y)} + \sin \varphi v_k^{(x)}
\end{pmatrix}
\end{align}
for $k=1,\dots,N$. In the following sections we will use this vector representation and the differential equations to describe the system dynamics.\\

In the following, we treat separately the cases of  abnormal (Subsect.~\ref{sec:abnormalsols}) and normal (Subsect.~\ref{sec:normalsols}) extremals for the case $N=2$, for which we provide semi-analytic solutions. 




\subsection{Abnormal extremals for N=2}\label{sec:abnormalsols}
We now fix $N=2$ and consider abnormal extremals, i.e. extremals with $\chi_0=0$. Our goal is to provide general expressions for the detuning $\Delta(t)$ of the laser phase $\varphi(t)$ -- i.e. the control -- and then to provide explicit proof of non-existence and time optimality for {\bf Cases (i)} and {\bf (ii)}, respectively.\\

Using the change of coordinates introduced in Eq.~\eqref{description}, Equations~\eqref{ham100} and~\eqref{H=0} together give 
\begin{equation}\label{eqxy}
v_1^{(x)}+\sqrt{2}v_2^{(x)}=v_1^{(y)}+\sqrt{2}v_2^{(y)}=0.
\end{equation}
Taking the time derivatives, Eq.~\eqref{humongous} leads to $$\cos\varphi (v_1^{(z)}+2v_2^{(z)})=\sin\varphi (v_1^{(z)}+2v_2^{(z)})=0,$$ which implies $v_1^{(z)}+2v_2^{(z)}=0$. By taking again the time derivative we obtain 
\begin{equation}
    (v_1^{(y)}+2\sqrt2v_2^{(y)})\cos\varphi+(v_1^{(x)}+2\sqrt2v_2^{(x)})\sin\varphi=0,
\end{equation}
which, together with Eq.~\eqref{eqxy}, reads
\begin{equation}\label{eqxysincos}
    v_1^{(y)}\cos\varphi+v_1^{(x)}\sin\varphi
=0.\end{equation}
From Eq.~\eqref{humongous} we see that $\dot z_1 = 0$, so that there exists some constant $a\in\mathbb{R}$ such that $z_1(t)=a$ at each time $t$.\\
Furthermore, in the new coordinates we identify the conserved quantity 
\begin{equation}\label{conservedquantityxy}
    \big(v_k^{(x)}\big)^2+\big(v_k^{(y)}\big)^2.
\end{equation}
For $k=1$, together with Eq.~\eqref{eqxysincos} the latter gives
\begin{align}\label{v1}
\mathbf{v}_1 &= 
\begin{pmatrix}
b\cos \varphi \\
-b\sin \varphi \\
a
\end{pmatrix}
\end{align}
for some constant $b\in\mathbb{R}$. Finally, we compute
\begin{equation}
    -\dot\varphi\sin\varphi=\frac{\mathrm{d}}{\mathrm{d}t}(\cos\varphi)=\frac{\dot v_1^x}{b}=-\frac{v_1^z\sin\varphi}{b}=-\frac{a\sin\varphi}{b}, 
\end{equation}
i.e. \begin{equation}\label{detuningconstant}
    \Delta(t)=\dot\varphi(t)=\frac{a}{b}.
\end{equation} 
We conclude that abnormal extremals to Eq.~\eqref{QOC} for $N=2$ TLSs give rise to a detuning that is constant in time. Therefore, the control is
\begin{equation}
    \varphi(t)=\frac{a}{b}t+c, \quad\text{for $c\in\mathbb{R}$.}
\end{equation}

In the following, we analyze the abnormal extremals of \textbf{Case (i)} and \textbf{Case (ii)} separately.

\subsubsection*{\textbf{Case (i)}: Non-existence of abnormal solutions}
We recall the target manifold defined in Eq.~\eqref{torus}, corresponding to the physical model where two TLSs are to be steered from their ground states to their respective excited states (in minimal time). At time $t=0$, 
\begin{equation}
    a=v_1^{(z)}(0)=\mathrm{Im}\langle\chi_1(0)|\psi_1(0)\rangle\overset{\eqref{Imaginary}}{=}0,
\end{equation}
since $\ket{\psi_1(0)}=\ket{0}_1$ and $\sigma_z\ket{0}_1=\ket{0}_1$.  Equation~\eqref{detuningconstant} dictates that the detuning $\Delta(t)$ is zero at each time $t$. To be more precise, this condition is to be understood as the detuning vanishing almost everywhere \cite{AdamsFournier, Brezis}, so that its primitive $\varphi(t)$ is a \textit{piecewise constant} function of time. We claim, however, that piece-wise extremals actually cannot give rise to an analytic solution to Eq.~\eqref{QOC} for this model, i.e. they do not exist. We show here why: 

Consider the evolution operator $U(t)$ as defined in \cite{Nielsen2010}.
Say that $\varphi(t)=\varphi_j$ for $t\in(t_{j-1},t_j]$ and $j=0,\dots,M$, where $M>0$ is some fixed integer and $t_j-t_{j-1}=T/M$ for each index $j$. We remark that $t_0=0$ and $t_M=T$. We can write the temporal operator as \begin{equation}\label{exp}
    U_k(T)=U_k(T,t_{m-1})U_k(t_{m-1},t_{m-2})\dots U_k(t_1,0)
\end{equation}
and \begin{align}\label{lastexp}
    U_k(t_j,t_{j-1}) &= \exp\left(-i H_k(t_j)\cdot \left( t_j - t_{j-1} \right)\right) \nonumber \\
    &= \exp\left(-i\frac{T}{M} H_k(t_j)  \right) \nonumber \\
    &= \exp \left(
    -i\frac{T}{M}\frac{\sqrt{k}}{2}
    \begin{pmatrix}
        0 & e^{i\varphi_j} \\
        e^{-i\varphi_j} & 0
    \end{pmatrix} \right).
\end{align}
From here, it is easy to compute the matrix exponential in Eq. \eqref{lastexp} and obtain
\begin{equation}U_k(t_j,t_{j-1})=
\begin{pmatrix}
\cos\left(\frac{\sqrt{k}T\varphi_j}{2M}\right) & -i \sin\left(\frac{\sqrt{k}T\varphi_j}{2M}\right) \\
i \sin\left(\frac{\sqrt{k}T\varphi_j}{2M}\right) & \cos\left(\frac{\sqrt{k}T\varphi_j}{2M}\right)
\end{pmatrix}.
\end{equation}
Now, we can rewrite \eqref{exp} as
\begin{equation}
    U_k(T)=\begin{pmatrix}
u_k & w^*_k \\
w_k & u_k
\end{pmatrix}
\end{equation}

with 
\begin{equation}
    \begin{cases}
        u_k = \cos\left(\frac{\sqrt{k}T}{2M}\left(\sum_{j=0}^{M}(-1)^{M-j}\varphi_j\right)\right) \\
        w_k = i \sin\left(\frac{\sqrt{k}T}{2M}\left(\sum_{j=0}^{M}\varphi_j\right)\right)
    \end{cases}.
\end{equation}
Hence, if we consider the evolution of the system at time $t=T$, we see that 
\begin{equation}
\begin{aligned}
    \ket{\psi_1(T)} &= U_1(T) \ket{0}
    = \begin{pmatrix}
    u_1 \\
    w_1
    \end{pmatrix}
\end{aligned}
\end{equation}
and
\begin{equation}
\begin{aligned}
    \ket{\psi_2(T)} &= U_2(T) \ket{0}
    = \begin{pmatrix}
    u_2 \\
    w_2
    \end{pmatrix}
\end{aligned}.
\end{equation}
Since we want $\ket{\psi_k(T)}$ to be a multiple of $\ket{1}_k$ up to some global phase, we need $\frac{T}{2M} \left(\sum_{j=0}^{M}(-1)^{M-j}\varphi_j \right)$ and $\frac{\sqrt{2}T}{2M} \left(\sum_{j=0}^{M}(-1)^{M-j}\varphi_j \right)$ both to be integer multiples of $\pi$, which is not possible since they differ by a factor of $\sqrt{2}$ which is irrational. This concludes the proof of non-existence of abnormal solutions in this case.

\subsubsection*{\textbf{Case (ii)}: Non-optimality of abnormal solutions}

We consider the target manifold described by Eq.~\eqref{1dimtarget}, which corresponds to the implementation of a CZ gate on two TLSs, up to single-TLS unitary gates. In the following, we prove that abnormal candidate extremals, when existent, are always slower than the time-optimal solution found in \cite{Sven}.

We recall that for the abnormal solutions considered here the detuning is constant, see   Eq.~\eqref{detuningconstant}. We then re-express the Hamiltonians Eq.~\eqref{hk2} as a function of the detuning \cite{Pichler, Jandura2024}
\begin{equation}\label{hamdet}
    H_k = \Delta\ket{1}{}_k{}_k\bra{1}+\frac{\sqrt{k}}{2}\sigma_x.
\end{equation}
Hence, for $\Delta$ constant, 
\begin{equation}\label{long}
\begin{aligned}
\exp(-i H_k T)
&= e^{-i\Delta T/2}
\Bigg[
    \cos\!\left(
        \tfrac{1}{2}\sqrt{k + \Delta^2}\, T
    \right) I
\\[4pt]
&\quad - i \sin\!\left(
        \tfrac{1}{2}\sqrt{k + \Delta^2}\, T
    \right)
    \frac{
        \Delta \sigma_z + \sqrt{k}\,\sigma_x
    }{
        \sqrt{k + \Delta^2}
    }
\Bigg]
\end{aligned}
\end{equation}
for $k=1$ and $2$. In order for the initial state $\ket{0}_k$ to return to $\ket{0}_k$ at time $t=T$ (i.e., at the end of the pulse) up to a phase, we obtain
\begin{equation}
    T = \frac{2\pi l}{\sqrt{1+\Delta^2}}= \frac{2\pi l'}{\sqrt{2+\Delta^2}}
\end{equation}
for $l,l'\ge1$ integers. From Eq.~\eqref{long}, it follows that
\begin{equation}
    \Delta= \pm \sqrt{\frac{2l^2-l'^2}{l'^2-l^2}}
\end{equation}
and
\begin{equation}
    T = 2\pi\sqrt{l'^2-l^2}.
\end{equation}
Since the minimal distance between two square number is $3=2^2-1^2$, we see that $T\ge2\sqrt{3}\pi\approx10.88$, which is more than the optimal time $T=7.612$ found in \cite{Sven}. We conclude that any abnormal extremals in \textbf{Case (ii)} are slower than the solution found in \cite{Sven} and hence are not time optimal.

\begin{figure*}[hbtp]
    \centering

    \begin{subfigure}{0.49\textwidth}
        \centering
        \caption*{{\large\textbf{(a)}} $\Delta_0=1.26$, $V_0=-1.17$}
        \includegraphics[width=\textwidth]{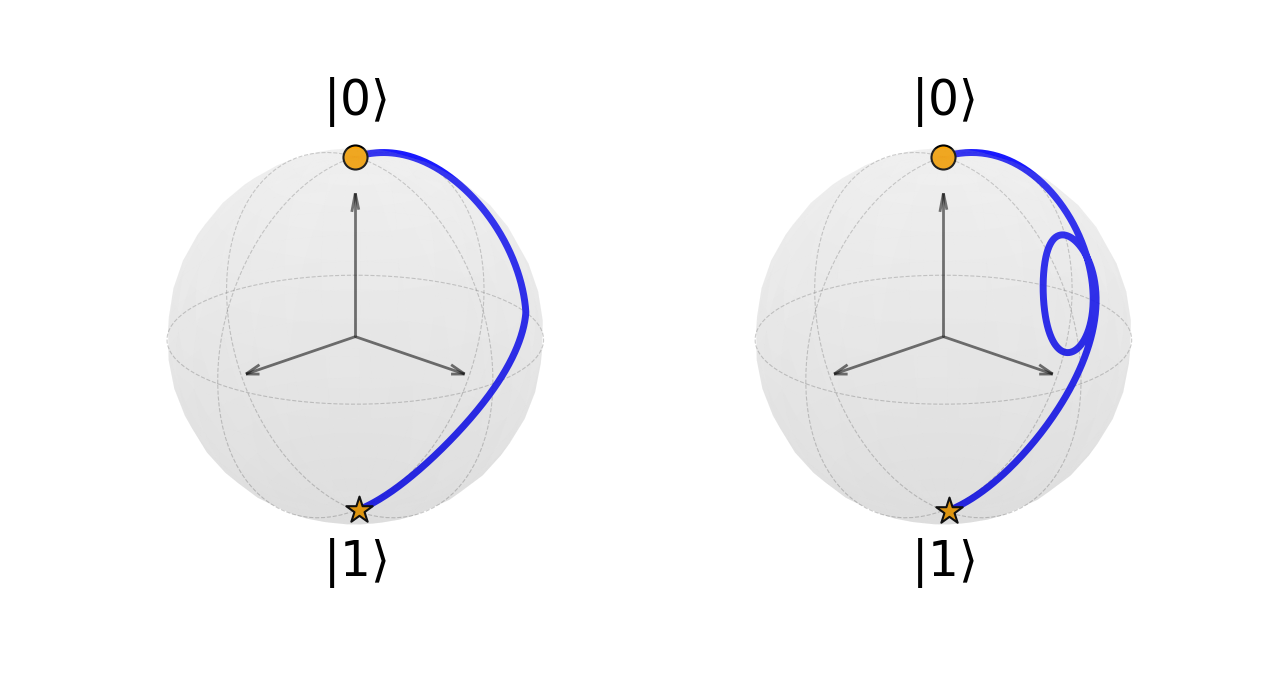}
    \end{subfigure}
    \hfill
    \begin{subfigure}{0.49\textwidth}
        \centering
        \caption*{{\large\textbf{(b)}} $\Delta_+=0.67$, $\Delta_-=-0.84$, $V_0=-0.39$}
        \includegraphics[width=\textwidth]{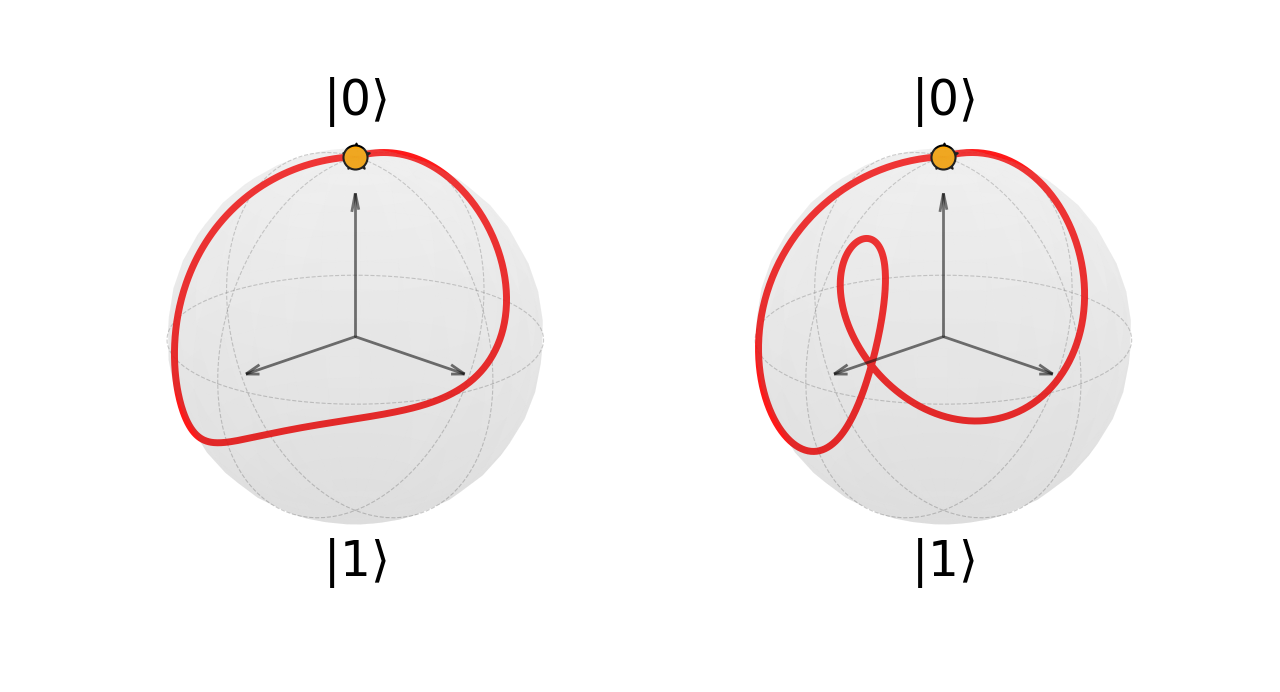}
    \end{subfigure}

    \vspace{-2em}

    \begin{subfigure}{0.49\textwidth}
        \centering
        \caption*{{\large\textbf{(c)}} $\Delta_0=0.77$, $V_0=0.02$}
        \includegraphics[width=\textwidth]{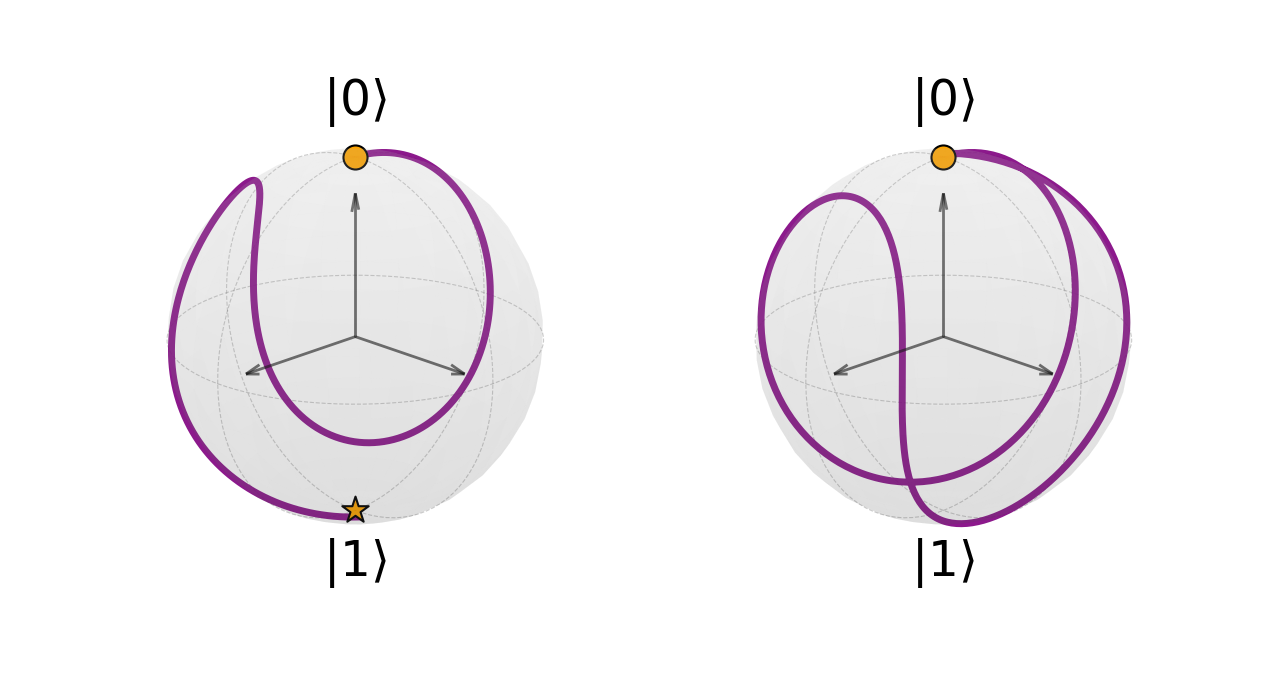}
    \end{subfigure}
    \hfill
    \begin{subfigure}{0.49\textwidth}
        \centering
        \caption*{{\large\textbf{(d)}} $\Delta_+=3.07$, $\Delta_-=-0.13$, $V_0=-1.00$}
        \includegraphics[width=\textwidth]{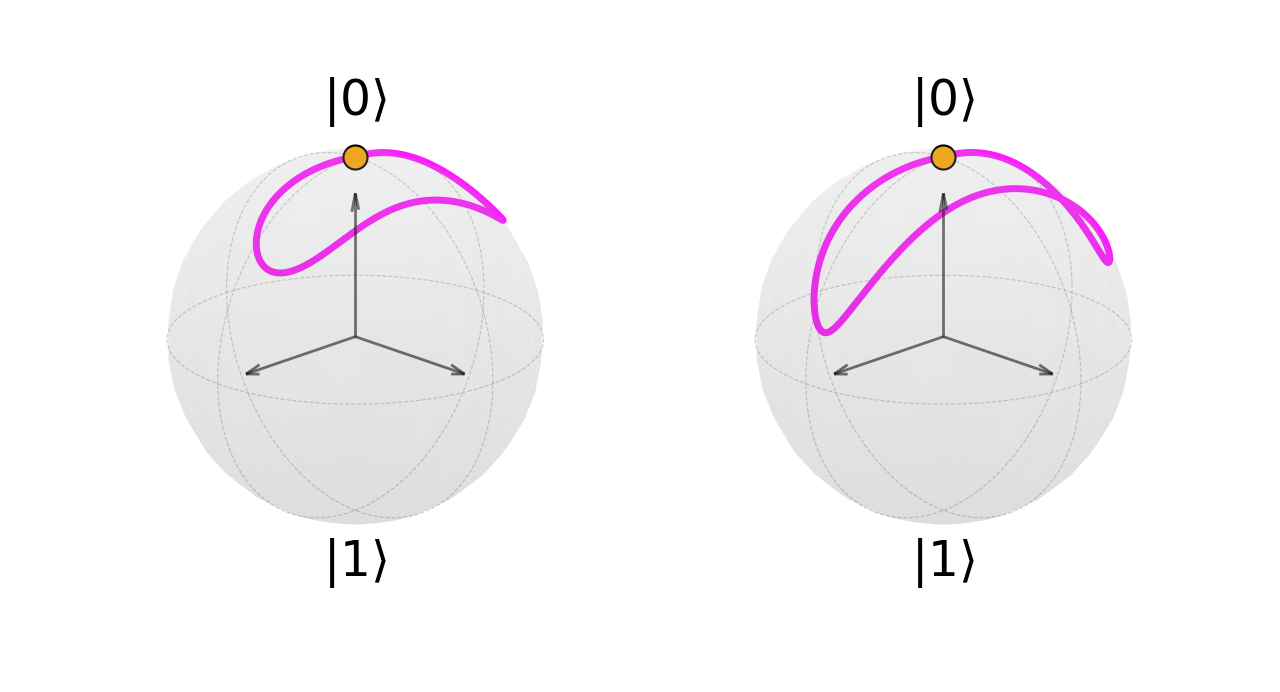}
    \end{subfigure}

    \vspace{-2em}

    \begin{subfigure}{0.49\textwidth}
        \centering
        \caption*{{\large\textbf{(e)}} $\Delta_0=1.01$, $V_0=-0.15$}
        \includegraphics[width=\textwidth]{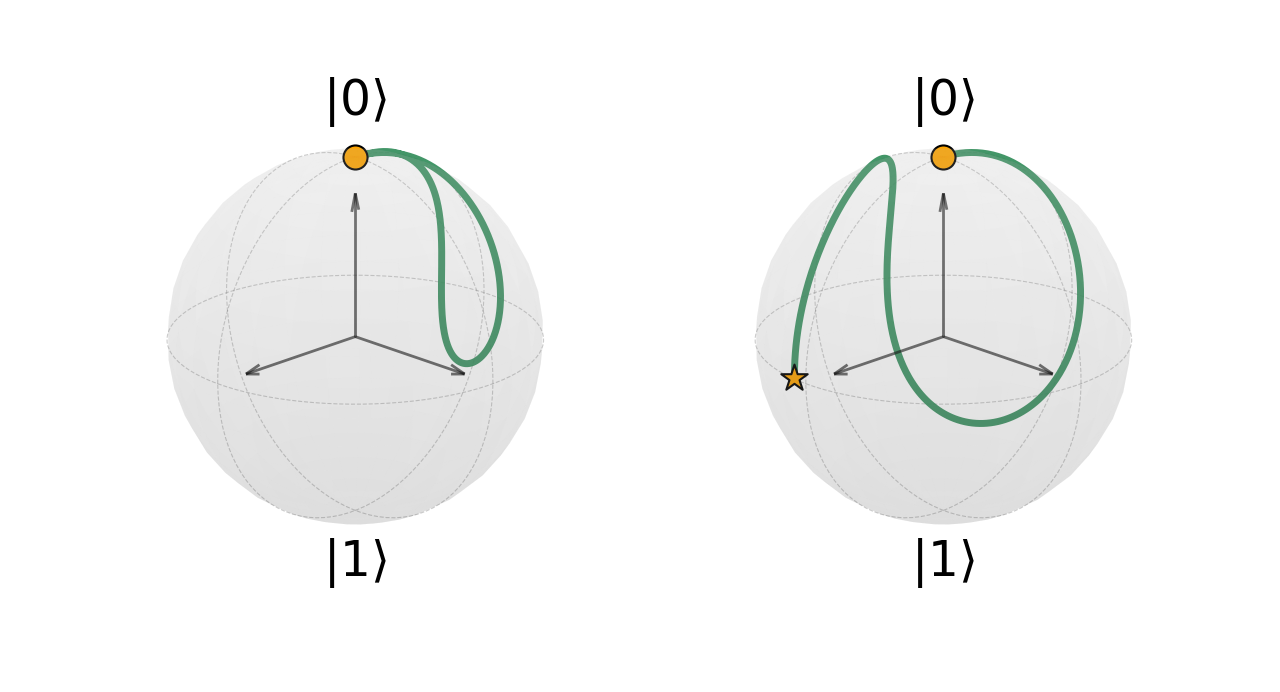}
    \end{subfigure}
    \hfill
    \begin{subfigure}{0.49\textwidth}
        \centering
        \caption*{{\large\textbf{(f)}} $\Delta_+=6.86$, $\Delta_-=-1.86$, $V_0=-35.02$}
        \includegraphics[width=\textwidth]{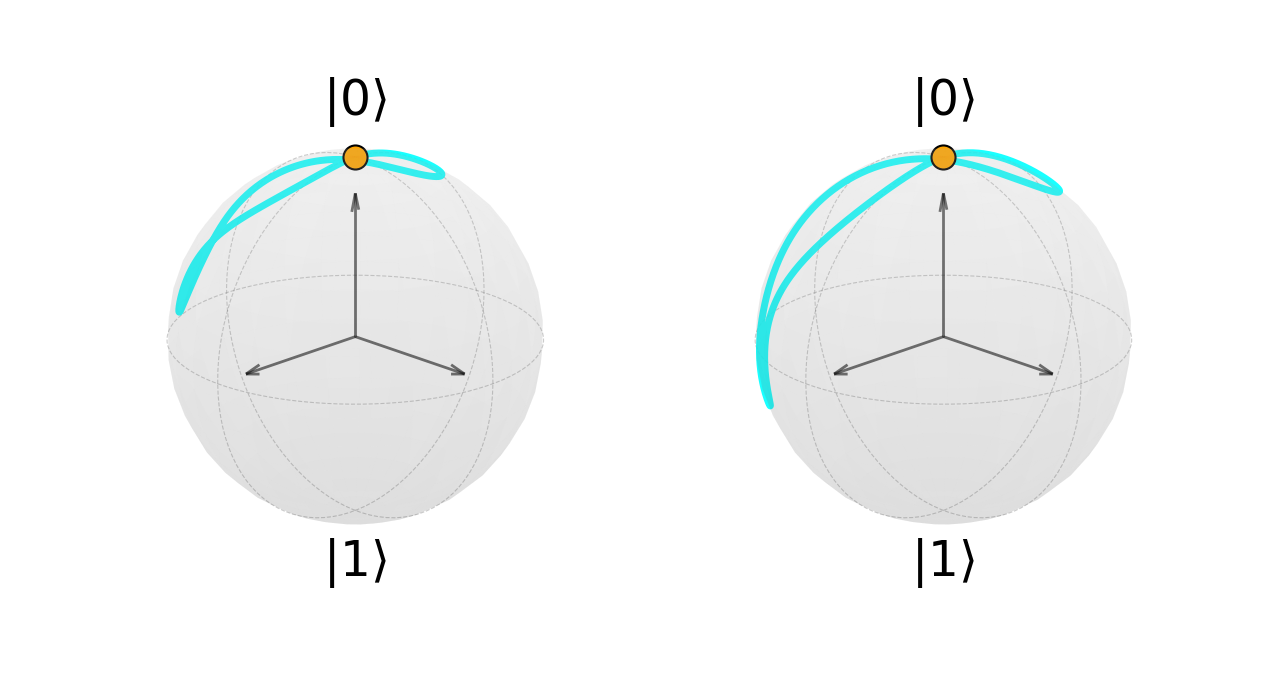}
    \end{subfigure}

    \vspace{-2em}

    \begin{subfigure}{0.49\textwidth}
        \centering
        \caption*{{\large\textbf{(g)}} $1.26\leq\Delta_0\leq1.50$, $V_0=-1.17$}
        \includegraphics[width=\textwidth]{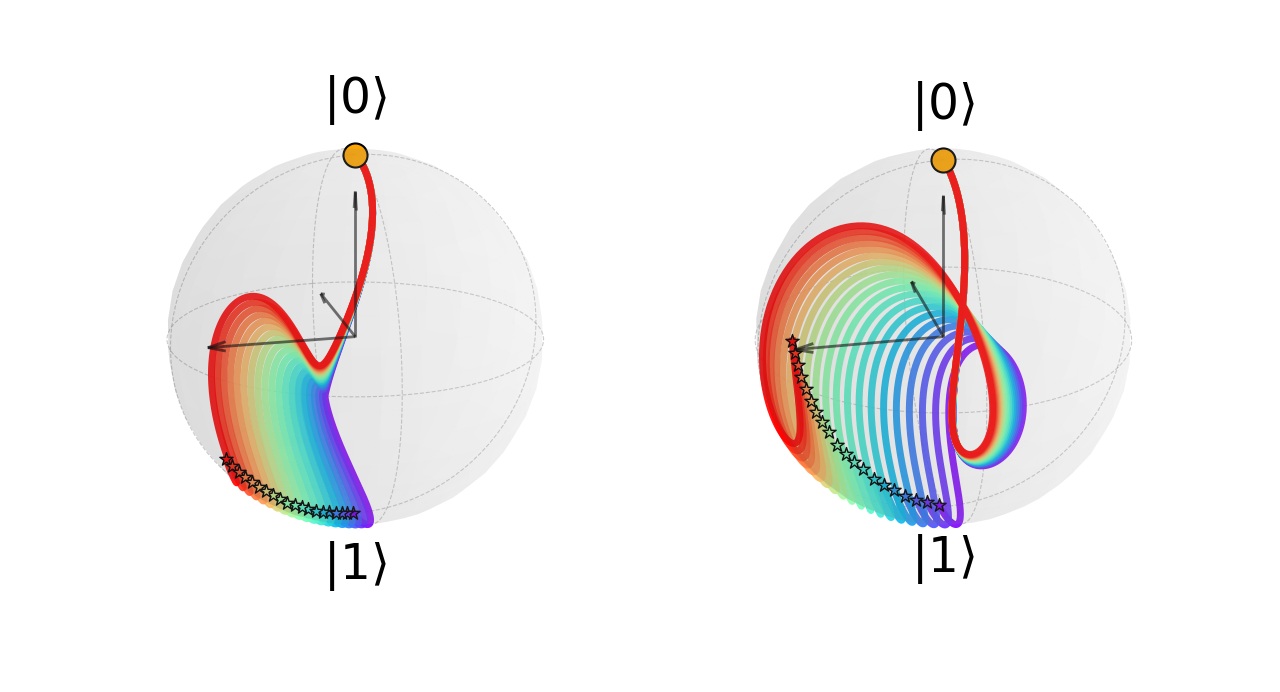}
    \end{subfigure}
    \hfill
    \begin{subfigure}{0.49\textwidth}
        \centering
        \caption*{{\large\textbf{(h)}} $\Delta_0 = 1.26$, $-1.50\leq V_0\leq-0.8$}
        \includegraphics[width=\textwidth]{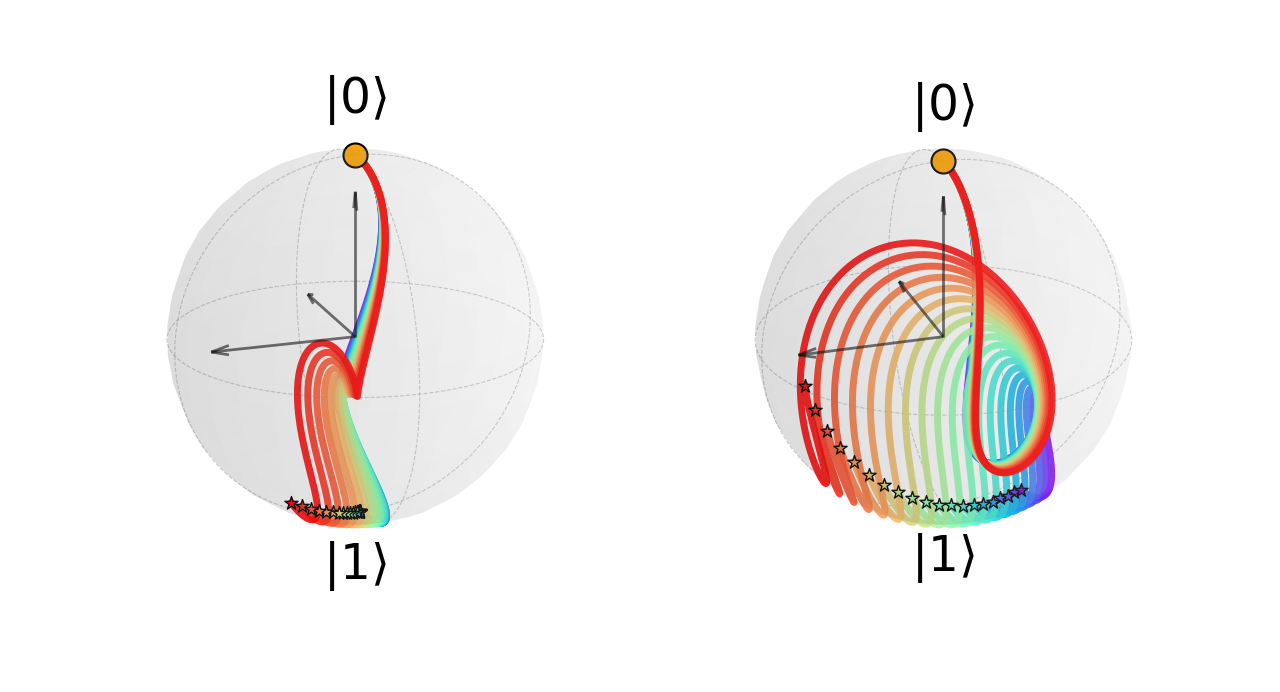}
    \end{subfigure}

    \caption{
Trajectories on the Bloch sphere corresponding to different normal extremals obtained with the PMP. 
Panels (a) and (b) show the trajectories associated with \textbf{Case (i)} and \textbf{Case (ii)}, respectively. 
Panel (c) displays an extremal that drives the first TLS from the ground to the excited state while keeping the second TLS in its ground state at the final time. 
In contrast to the CZ gate of \textbf{Case (ii)}, which introduces a phase $\pi$ on the second TLS, panels (d) and (f) show state-to-state transfers generating smaller conditional phase shifts of $\pi/3$ and $\pi/50$, respectively. 
Panel (e) illustrates an extremal in which the second TLS is steered to the equator of the Bloch sphere while the first TLS is driven back to its ground state. 
Panels (g) and (h) display trajectories obtained by varying the parameters $\Delta_0$ and $V_0$, respectively, highlighting the effect of these parameters on the extremal evolution. 
The starting point of each trajectory is marked by an orange dot, and the final point is indicated by a star; if the final point coincides with the starting point, only the dot is visible.
}
    \label{fig:trajectories}
\end{figure*}

\subsection{Normal extremals}\label{sec:normalsols}

At this point in the analysis, we consider normal extremals, i.e. extremals for which $\chi_0=-1$.  We first provide  expressions for the laser detuning $\Delta(t)$ for a general number of TLSs $N$.  For the case $N=2$, we then prove that the detuning $\Delta(t)$ of the laser pulse $\varphi(t)$ obeys the same ODE (Eq.~\eqref{eqdiffpotential}) of a classic particle moving in a quartic potential. Finally, we consider separately \textbf{Cases (i)} and \textbf{(ii)} and numerically integrate the respective ODEs. We found perfect agreement with numerical solutions found by GRAPE. For specific choices of parameters, our results reproduce the results of Refs. \cite{Pichler} and \cite{Sven} for \textbf{Cases (i)} and \textbf{(ii)}, respectively. Furthermore, we illustrate in Fig.~\ref{fig:trajectories} a variety of trajectories stemming from different target states.

From Eq.~\eqref{ham100} we derive
\begin{equation}\label{=4} \small
\left( \sum_k \sqrt{k} \Im \langle \chi_k | \sigma_x | \psi_k \rangle \right)^2 
+ \left( \sum_k \sqrt{k} \Im \langle \chi_k | \sigma_y | \psi_k \rangle \right)^2 = 4.
\end{equation}
This leads to a simplification of Eqs.~\eqref{cos} and~\eqref{sin}:
\begin{equation}\label{cos1}
    \cos \varphi = \frac{1}{2}\sum_k \sqrt{k} \Im \langle \chi_k | \sigma_x | \psi_k \rangle
\end{equation}
and 
\begin{equation}\label{sin1}
    \sin \varphi = -\frac{1}{2}\sum_k \sqrt{k} \Im \langle \chi_k | \sigma_y | \psi_k \rangle.
\end{equation}
Recall Eqs.~\eqref{description} and~\eqref{humongous}.
In this writing, the expression of the laser pulse $\varphi$ is 
\begin{equation} \label{trigon}
\begin{aligned}
    \cos\varphi &= \frac{1}{2}\sum_k\sqrt{k}v_k^{(x)}, \\
    \sin\varphi &= -\frac{1}{2}\sum_k\sqrt{k}v_k^{(y)}.
\end{aligned}
\end{equation}
Hereby, Eq.~\eqref{humongous} gives $3N$ coupled quadratic differential equations for the $v_k$.
In these coordinates, it is easy to recognize few conserved quantities:
\begin{itemize}
    \item The radii 
    \begin{equation}\label{rad}
        r_k = ||\mathbf{v}_k|| = \sqrt{\left(v_k^{(x)}\right)^2+\left(v_k^{(y)}\right)^2+\left(v_k^{(z)}\right)^2}
    \end{equation} are constant throughout the evolution of the system due to Eq.~\eqref{humongous}.
    \item From Eq.~\eqref{=4} we derive
    \begin{equation}\label{=4pt2}
        \left(\sum_k\sqrt{k}v_k^{(x)} \right)^2+\left(\sum_k\sqrt{k}v_k^{(y)} \right)^2 = 4.
    \end{equation}

    \item The quantity
\begin{equation}\label{C}
    C = \sum_k v_k^{(z)}
\end{equation} is constant due to Eqs.~\eqref{humongous} and~\eqref{trigon}.
    
\end{itemize}

Recalling the expression for the laser detuning as $\Delta(t)~
=~\dot{\varphi}(t)$, we obtain
\begin{align*}
    \Delta\cos\varphi &= \frac{d}{dt}\sin\varphi
    = -\frac{1}{2}\sum_k\sqrt{k}\dot v_k^{(y)} \\
    &= \frac{\cos\varphi}{2}\sum_kkv_k^{(z)}.
\end{align*}
Thus,
\begin{equation}\label{delta}
    \Delta = \frac{1}{2}\sum_kkv_k^{(z)}.
\end{equation}

So far we have reformulated the problem relative to the model describing the Rydberg blockade dynamics by means of $N$ of TLSs in Eq.~\eqref{TLS}. The PMP has provided interesting insights relative to the dynamics of the system and has endowed us with costates and conserved quantities that we need to take into account. Finally, we have proved that the laser detuning $\Delta=\dot{\varphi}$ can be expressed through the new chosen coordinates.
\subsection*{Exact results for $N=2$ TLS}

In the remainder of this work, we fix $N=2$.
Let us parametrize the vectors $\mathbf{v}_1$ and $\mathbf{v}_2$ by
\begin{equation}\label{parametrization}
\mathbf{v}_k =
\begin{pmatrix}
\sqrt{r_k^2-\xi_k^2}\cos\xi_k \\
\sqrt{r_k^2-\xi_k^2}\sin\xi_k \\
z_k
\end{pmatrix},
\end{equation}
with the new variables $\xi_1,\xi_2,z_1,z_2$.
From Eq. \eqref{=4pt2},
\begin{equation*} \small
\begin{aligned}
4 &= \left( \sqrt{r_1^2 - z_1^2} \cos \xi_1 + \sqrt{2} \sqrt{r_2^2 - z_2^2} \cos \xi_2 \right)^2 \nonumber \\
&\quad + \left( \sqrt{r_1^2 - z_1^2} \sin \xi_1 + \sqrt{2} \sqrt{r_2^2 - z_2^2} \sin \xi_2 \right)^2 \nonumber \\
&= r_1^2 - z_1^2 + 2(r_2^2 - z_2^2) + \sqrt{8} \sqrt{r_1^2 - z_1^2} \sqrt{r_2^2 - z_2^2} \cos (\xi_1 - \xi_2).
\end{aligned}
\end{equation*}
and thus
\begin{equation}\label{aaaq} \small
\sqrt{8} \sqrt{r_1^2 - z_1^2} \sqrt{r_2^2 - z_2^2} \cos (\xi_1 - \xi_2) = 4 - (r_1^2 - z_1^2) - 2(r_2^2 - z_2^2).
\end{equation}

\begin{figure*}[t]
    \centering
    \begin{tikzpicture}
        \node[anchor=south west, inner sep=0] (image) 
            {\includegraphics[width=\textwidth]{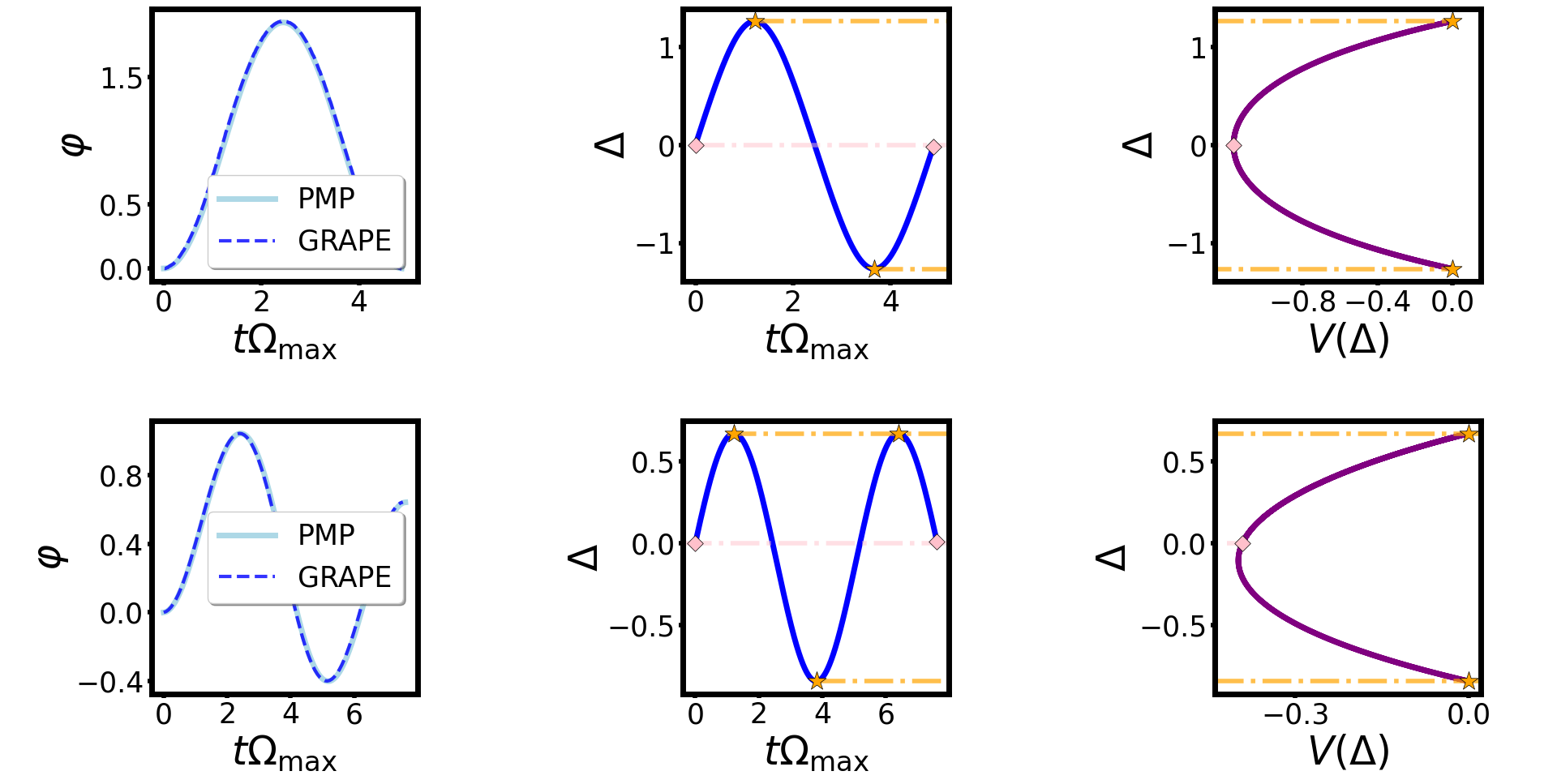}};
        
        \begin{scope}[x={(image.south east)}, y={(image.north west)}]
            \node[anchor=north west, font=\bfseries\large] at (0.02,0.98) {(a)};

            \node[anchor=north west, font=\bfseries\large] at (0.02,0.49) {(b)};

        \end{scope}
    \end{tikzpicture}
    \caption{Depicted are \textbf{Case (i)} and \textbf{Case (ii)}, shown respectively in panels (a) and (b). 
    The first column displays the laser pulse $\varphi(t)$, the second one the detuning $\Delta(t)$, both plotted as functions of time, 
    while the third column shows the potential $V(\Delta)$ as a function of the detuning. All quantities are normalized by $\Omega_{\text{max}}$. 
    The starred points mark the inversion points of the motion inside the potential, whereas the pink points indicate the starting and final points of the detuning which correspond to the starting and final point of the motion inside the potential. 
    The dashed blue curves represent the numerical GRAPE solutions, while the solid light-blue curves correspond to the semi-analytic results.}
    \label{fig:combined}
\end{figure*}

By taking the time derivative of $z_1$,
\begin{equation} \small
\begin{aligned} 
\dot{z}_1 &= \cos \varphi \, v_1^{(y)} + \sin \varphi \, v_1^{(x)} \\
&= \sqrt{r_1^2 - z_1^2} (\cos \varphi \sin \xi_1 + \sin \varphi \cos \xi_1) \\
&= \frac{1}{2} \sqrt{r_1^2 - z_1^2} \left( \sqrt{r_1^2 - z_1^2} \cos \xi_1 
+ \sqrt{2} \sqrt{r_2^2 - z_2^2} \cos \xi_2 \right) \sin \xi_1 \\
&\quad - \frac{1}{2} \sqrt{r_1^2 - z_1^2} \left( \sqrt{r_1^2 - z_1^2} \sin \xi_1 
+ \sqrt{2} \sqrt{r_2^2 - z_2^2} \sin \xi_2 \right) \cos \xi_1 \\
&= \frac{1}{\sqrt{2}} \sqrt{r_1^2 - z_1^2} \sqrt{r_2^2 - z_2^2} \sin (\xi_1 - \xi_2)
\end{aligned}
\end{equation}
and we conclude that 
\begin{equation}\label{svhwc}
    \sqrt{8}\sqrt{r_1^2-z_1^2}\sqrt{r_2^2-z_2^2}\sin(\xi_1-\xi_2)=4\dot{z}_1.
\end{equation}
By adding the squares of Eqs.~\eqref{aaaq} and~\eqref{svhwc}, we obtain
\begin{equation}\small\label{wcjkw}
    8(r_1^2-z_1^2)(r_2^2-z_2^2)=16\dot{z}_1^2+[4-(r_1^2-z_1^2)-2(r_2^2-z_1^2)]^2.
\end{equation}
From Eqs. \eqref{C} and \eqref{delta}, we obtain $z_1 + 2 z_2 = 2 \Delta$ and $z_1 + z_2 = C$. Hence,
\begin{align}
    z_1 &= -2\Delta + 2C, \\
    z_2 &= 2\Delta - C. 
\end{align}
Substituting this into Eq.~\eqref{wcjkw} and using that $\dot{C} = 0$ (from Eq.~\eqref{C}) gives
\begin{equation}\label{eqdiffpotential}
    \frac{1}{2}\dot{\Delta}^2 + V(\Delta) = 0. 
\end{equation}
where
\begin{equation}\label{generalpotential}
    V(\Delta) = c_4 \Delta^4 + c_3 \Delta^3 + c_2 \Delta^2 + c_1 \Delta + c_0
\end{equation}
with
\begin{align}
    c_4 &= \frac{1}{8}, \; 
    c_3 = 0,  \;
    c_2 = \frac{-2C^2 + r_1^2 - 2 r_2^2 + 12}{16}, \;
    c_1 = -C,\notag\\
    c_0 &= \frac{\left( 6 C^2 - r_1^2 - 2 r_2^2 + 4 \right)^2 - 8 \left( 4 C^2 - r_1^2 \right) \left( C^2 - r_2^2 \right)}{128}.\notag
\end{align}
From Eq.~\eqref{eqdiffpotential} we understand a surprising fact about extremals of the OC problem in Eq.~\eqref{QOC}: they admit a laser detuning $\Delta=\dot{\varphi}$ that has the same functional dependency of a particle moving in the quartic potential of Eq.~\eqref{generalpotential}. This is one of the main results of this work. \\

In the following, we focus on \textbf{Cases (i)} and \textbf{(ii)}, and we numerically solve the ODE in Eq.~\eqref{eqdiffpotential}, which governs the evolution of the detuning. We stress, however, that these two examples represent only a subset of the many possible state-to-state transfers that can be constructed within the semi-analytic framework developed in this work. In Fig.~\ref{fig:trajectories}, we display the trajectories corresponding to several different transfers obtained using the same computational procedure adopted in the next sections for\textbf{ Case (i)} and \textbf{Case (ii)}.


\subsubsection*{\textbf{Case (i)} with $N=2$}\label{sec:subsubsec1}
We recall the target manifold defined in Eq.~\eqref{torus}, that corresponds to the physical model where two TLSs are to be steered from their ground states to their respective excited states (see Ref. \cite{Pichler}). In this case, by considering Eq.~\eqref{Imaginary}, it is easy to see that the conserved quantity $C$ in Eq.~\eqref{C} is zero, i.e.
\begin{equation}\label{C=0}
    C=v_1^{(z)} + v_2^{(z)}=0 \quad \text{for each time $t$.}
\end{equation}
Therefore, we rewrite the detuning as
\begin{equation}\label{detuningv1z}
    \Delta = -\frac{1}{2}v_1^{(z)}.
\end{equation}
The potential takes the form 
\begin{equation}\label{potential}
    V(\Delta)=c_4\Delta^4+c_2\Delta^2+c_0
\end{equation}
Finally, we observe that at times  $t=0$ and $t=T$, the $\sigma_z|\psi_k\rangle$ are multiples of $|\psi_k\rangle$; consequently, Eqs.~\eqref{Imaginary} and~\eqref{detuningv1z} imply that $\Delta(0)=\Delta(T)=0$. Hence, extremals respond to laser pulses whose detuning obeys to Eq.~\eqref{eqdiffpotential} and vanishes at the beginning and at the end of the pulse.\\
In order to obtain a normal extremal, it remains to choose the coefficients $c_0$ and $c_2$ in Eq.~\eqref{potential} so that the laser pulse $\varphi(t)=\int_0^t\Delta(s)\text{d}s+\varphi(0)$ steers the ground states $\ket{0}_1$ and $\ket{0}_2$ to their excited states $\ket{1}_1$ and $\ket{1}_2$, respectively.
It is useful to rewrite the potential with a different parametrization. Since $V_0:=V(0)<0$, the potential has exactly two real roots: $\Delta_0$ and $-\Delta_0$. Hence we can write it as \begin{equation}\label{potentialdelta0v0}
    V(\Delta)=(\Delta^2-\Delta_0^2)\left(\frac{1}{8}\Delta^2-\frac{V_0}{\Delta_0^2}\right).
\end{equation}

We now proceed with determining the parameters $\Delta_0$ and $V_0$ as follows. Without loss of generality, suppose that $\dot\Delta(0)>0$ (the opposite case $\dot\Delta(0)<0$ can be retrieved by means of the transformations $t\mapsto T-t$, $\varphi\mapsto2\pi-\varphi$ and $\Delta\mapsto-\Delta$: interestingly, time-optimal pulses are not unique, but exhibit a limited number of symmetries, with time reversal being one of them
\footnote{See \cite{Sven} for other symmetries relevant to the realization of general C-phase gates, including the CZ gate considered in the state-to-state transfer of \textbf{Case (ii)}.}). For some fixed pair $\Delta_0>0$ and $ V_0<0$, we numerically solve the differential equation
\begin{equation}
    \ddot\Delta(t) = - \frac{\mathrm{d}V(\Delta)}{\mathrm{d}\Delta}
\end{equation}
with $\Delta(0)=0$ and $\dot\Delta(0)=\sqrt{-2V_0}$ as initial conditions. We remark that the equation above can be obtained by taking the time derivative of Eq.~\eqref{eqdiffpotential}.

At first, we choose a tentative final time \(T\) such that the numerical solution satisfies \(\Delta(T)=0\). From the resulting detuning \(\Delta(t)\), we then compute the laser phase
$
\varphi(t) = \int_0^t \Delta(s)\,\mathrm{d}s+\varphi(0) ,
$
and insert it into the Hamiltonian in Eq.~\eqref{hk2}. This allows us to determine numerically the corresponding trajectories of the states \( |\psi_1(t)\rangle \) and \( |\psi_2(t)\rangle \).

 We are left to evaluate the fidelity function $\tilde F = \min_{\alpha,\beta}F$, with
\begin{equation}\label{fidelity}
\begin{aligned}
F = \frac{1}{6}\Bigl(
 &\bigl| e^{-i\alpha}{}_1\langle 1|\psi_1(T)\rangle
      + e^{-i\beta}{}_2\langle 1|\psi_2(T)\rangle \bigr|^2 \\[6pt]
 &\quad + |{}_1\langle 1|\psi_1(T)\rangle|^2
        + |{}_2\langle 1|\psi_2(T)\rangle|^2
\Bigr).
\end{aligned}
\end{equation}
We numerically optimize $\tilde F$ over $\Delta_0$ and $V_0$ using the BFGS algorithm \cite{SciML} and we confront the detuning obtained as a solution of Eq.~\eqref{eqdiffpotential} with the one obtained from the GRAPE algorithm.\\
In Fig.~\ref{fig:combined}, we display a dashed blue curve representing the GRAPE solution, a solid light-blue curve for the PMP extremal, and a purple curve for the corresponding potential.
The first upper panel (a) corresponds to the extremal of the OC problem with target manifold described by Eq.~\eqref{torus}, namely \textbf{Case (i)}. In this case, it can be seen that the trajectory performs one full oscillation in the potential in order to correctly steer the two TLSs from their ground state to their excited state in the optimal time $T = 4.875$, as found in~\cite{Pichler}. We observe an excellent agreement between the pulses found by GRAPE and by the PMP, confirming that the semi-analytic pulses derived in this chapter faithfully capture the time-optimal solutions.

\subsubsection*{\textbf{Case (ii)} with $N=2$}
We consider the target manifold defined in Eq.~\eqref{1dimtarget}, which characterizes the implementation of a CZ gate on two TLSs, modulo single-TLS unitary operations. In general, Eq.~\eqref{C=0} does not hold, therefore the potential is as in Eq.~\eqref{generalpotential}.
Similarly to \textbf{Case (i)} above, one can prove \cite{Jandura2024} that $\Delta(0)=\Delta(T)=0$. Furthermore, the potential admits two real roots $\Delta_+>0$ and $\Delta_-<0$ such that the potential can be expressed as
\begin{align}
    V(\Delta) = & \, (\Delta - \Delta_+)(\Delta - \Delta_-) \bigg( \frac{1}{8} \Delta^2  \notag \\
    & + \, \frac{1}{8} \big(\Delta_+ + \Delta_-\big) \Delta + \frac{V_0}{\Delta_+ \Delta_-} \bigg).
\end{align}
We determine the extremals by proceeding as for \textbf{Case (i)} above and we find the same results of \cite{Sven}. In particular, in Fig.~\ref{fig:combined} (b) we confront the extremal control $\varphi$ with the GRAPE solution in the first column, and we display in the second and third columns respectively the detuning versus its associated potential. This picture is interesting since it manifestly shows the differences with the curves plotted in Fig.~\ref{fig:combined} (a). The two branches of the potential are not symmetric with respect to each other, and consequently the detuning loses the symmetry between its maxima and minima. Finally, we remark that in this case the detuning performs one and a half oscillation inside the potential.

\subsubsection*{\textbf{General Cases}}

Notably, we remark that every time-optimal state-to-state transfer devised in the Rydberg blockade regime is driven by a laser pulse $\varphi$ whose detuning $\Delta$ is a solution of Eq.~\eqref{eqdiffpotential}. Hereby, \textbf{Case (i)} and \textbf{Case (ii)} only represent two examples of the transfers achievable in this paradigm. To provide further examples, we plot in Fig.~\ref{fig:trajectories} the trajectories of some normal extremals found by our semi-analytic approach. In particular, in panel (a) and (b) respectively we plot the trajectories relative to \textbf{Cases (i)} and \textbf{(ii)}. In panels (c) and (e) we show trajectories representing the steering of ground states to different combinations of the effective ground and excited states.  In  panels (d) and (f) we provide the time evolutions of trajectories for the realization of Controlled-Phase (C-Phase) gates \cite{Sven} alternative to the aforementioned CZ-gate of \textbf{Case (ii)}. Finally, in panels (g) and (h) we showcase the behavior of the trajectories when the detuning amplitude $\Delta_0$ (see Eq.~\eqref{potentialdelta0v0})  and the potential minimum $V_0$ respectively vary.

\section{Conclusion}
In this work we derive a semi-analytic approach to the problem of time-optimal state-to-state transfer for an ensemble of independent, generally non-equivalent effective two-level subsystems using the Pontryagin Maximum Principle. This situation is relevant for the case of optimal laser pulses globally addressing $N$ Rydberg atoms in the blockade regime. We show how the PMP provides meaningful insights to the optimization problem and then we focus on the low-dimensional case $N=2$. One main result of this paper is that, for $N=2$, the detuning of an optimal \textit{normal} solution obeys the same ODE of the position of a particle immersed in a quartic potential. We leverage this expression and explicitly solve for two different state-to-state transfers with applications in quantum computing. However, our approach is rather general and shows that the Pontryagin Maximum Principle can be applied to certain multi-qubit problems of considerable importance in quantum science and technology. In the future, it will be interesting to investigate whether similar techniques and analytical insights can be applied to more general multi-qubit entangling operations.

\section*{Acknowledgments}
This research has received funding from European Union’s Horizon Europe Research and Innovation Programme under the Marie Skłodowska-Curie GA number 10120240 (MLQ), the HORIZON-CL4-2021-DIGITAL-EMERGING-01-30 via the project 101070144 (EuRyQa)  and from the French National Research Agency under the Investments of the Future Program projects ANR-21-ESRE-0032 (aQCess), ANR-22-CE47-0013-02 (CLIMAQS), ANR-17-EURE-0024 (QMat), and ANR-22-CMAS-0001 France 2030 (QuanTEdu-France).



\appendix 
\section{}
\label{sec:appendixA}
The fact that the state manifold is complex demands for a bridging with the real set-up formulation of the PMP~\cite{DAlessandro}. First, we replace each inner product \( \langle x, y \rangle \) in the formulation of the PMP by the real part of the inner product between quantum states 
\[
\operatorname{Re}(\langle \phi | \psi \rangle) = \big\langle \operatorname{Re}(|\psi \rangle), \operatorname{Re}(|\phi \rangle) \big\rangle + \big\langle \operatorname{Im}(|\psi \rangle), \operatorname{Im}(|\phi \rangle) \big\rangle.
\]
Second, we substitute the gradients by Wirtinger derivatives. The Wirtinger derivative of a real function \( f(|\psi \rangle) \) depending on a complex state \( |\psi \rangle \) is defined as
\[
\frac{\partial f}{\partial |\psi \rangle} = \frac{\partial f}{\partial \operatorname{Re}(|\psi \rangle)} - i \frac{\partial f}{\partial \operatorname{Im}(|\psi \rangle)}
\]
where \( \operatorname{Re}(|\psi \rangle) \) and \( \operatorname{Im}(|\psi \rangle) \) are treated as independent variables. The Wirtinger derivative is the natural extension of the derivative to functions with complex inputs and real outputs, as it satisfies
\[
f(|\psi \rangle + \varepsilon |\phi \rangle) = f(|\psi \rangle) + \varepsilon \operatorname{Re} \left( \left\langle \frac{\partial f}{\partial |\psi \rangle} \Big| \phi \right\rangle \right) + \mathcal{O}(\varepsilon^2)
\] as $\varepsilon\to0$.




\bibliography{bibliografia}

\end{document}